\catcode`\@=11
\newif\if@fewtab\@fewtabtrue

{\count255=\time\divide\count255 by 60
\xdef\hourmin{\number\count255}
\multiply\count255 by-60\advance\count255 by\time
\xdef\hourmin{\hourmin:\ifnum\count255<10 0\fi\the\count255}}
\def\ps@draft{\let\@mkboth\@gobbletwo
    \def\@oddhead{}
    \def\@oddfoot
       {\hbox to 7 cm{$\scriptstyle Draft\ version:\ \draftdate$
       \hfil}\hskip -7cm\hfil\rm\thepage \hfil}
    \def\@evenhead{}\let\@evenfoot\@oddfoot}

\def\ceqno{\global\@fewtabfalse
    \ifcase\@eqcnt \def\@tempa{& & &}\or \def\@tempa{& &}
      \or \def\@tempa{&}
      \or\def\@tempa{}\fi\@tempa
{\rm(\theequation)}}

\def\aeqno#1{\global\@fewtabfalse
    \ifcase\@eqcnt \def\@tempa{& & &}\or \def\@tempa{& &}
      \or \def\@tempa{&}
      \or\def\@tempa{}\fi\@tempa
{\rm(\theequation,#1)}}

\def\label#1{\ifnum\draftcontrol=1
 \global\def\draftnote{$\scriptstyle #1$}\fi
 \@bsphack\if@filesw {\let\thepage\relax
   \def\protect{\noexpand\noexpand\noexpand}%
\xdef\@gtempa{\write\@auxout{\string
      \newlabel{#1}{{\@currentlabel}{\thepage}}}}}\@gtempa
   \if@nobreak \ifvmode\nobreak\fi\fi\fi
  \@esphack}

\def\alabel#1#2{\label{#1}\global\@fewtabfalse
    \ifcase\@eqcnt \def\@tempa{& & &}\or \def\@tempa{& &}
      \or \def\@tempa{&}
      \or\def\@tempa{}\fi\@tempa
{\hbox to 3cm{\phantom{\rm(\theequation,#2)}
\draftnote \hfil}\hskip -3cm {\rm(\theequation,#2)}}}

\def\clabel#1{\label{#1}\global\@fewtabfalse
    \ifcase\@eqcnt \def\@tempa{& & &}\or \def\@tempa{& &}
      \or \def\@tempa{&}
      \or\def\@tempa{}\fi\@tempa
{\hbox to 3cm{\phantom{\rm(\theequation)}
\draftnote \hfil}\hskip -3cm{\rm(\theequation)}}}

\def\eqnarray{\def\draftnote{{}}\global\@fewtabtrue
\stepcounter{equation}\let\@currentlabel=\theequation
\global\@eqnswtrue
\global\@eqcnt\z@\tabskip\@centering\let\\=\@eqncr
$$\halign to \displaywidth\bgroup\@eqnsel\hskip\@centering\@eqcnt\z@
  $\displaystyle\tabskip\z@{##}$&\global\@eqcnt\@ne
  \hskip 1\arraycolsep \hfil${##}$\hfil
  &\global\@eqcnt\tw@ \hskip 1\arraycolsep
$\displaystyle\tabskip\z@{##}$
\hfil  \tabskip\@centering&\global\@eqcnt\thr@@\llap{##}\tabskip\z@
\cr}

\def\endeqnarray{\@@eqncr\egroup
      \global\advance\c@equation\m@ne$$\global\@ignoretrue}

\def\@eqnnum{\hbox to 3cm{\phantom{\rm(\theequation)} \draftnote
                         \hfil}\hskip -3cm {\rm(\theequation)}}

\def\@@eqncr{\let\@tempa\relax
    \ifcase\@eqcnt \def\@tempa{& & &}\or \def\@tempa{& &}
      \or \def\@tempa{&}
      \or\def\@tempa{}
\fi\@tempa
\if@eqnsw
\if@fewtab\@eqnnum\fi
\stepcounter{equation}\fi\global
\@eqnswtrue\global\@eqcnt\z@\global\@fewtabtrue\cr}

\def\draftcite#1{\ifnum\draftcontrol=1#1\else{}\fi}

\def\@lbibitem[#1]#2{\item{}\hskip -3cm \hbox to 2cm
{\hfil$\scriptstyle\draftcite{#2}$}\hskip
1cm[\@biblabel{#1}]\if@filesw
     {\def\protect##1{\string ##1\space}\immediate
      \write\@auxout{\string\bibcite{#2}{#1}}}\fi\ignorespaces}

\def\@bibitem#1{\item\hskip -3cm \hbox to 2cm
{\hfil $\scriptstyle\draftcite{#1}$}\hskip 1cm
\if@filesw \immediate\write\@auxout
       {\string\bibcite{#1}{\the\value{\@listctr}}}\fi\ignorespaces}

\def\nsection#1{\section{#1}\setcounter{equation}{0}}

\font\tendl=msbm10  scaled \magstep1
\font\sevendl=msbm7 scaled \magstep1
\font\fivedl=msbm5 scaled \magstep1
\font\tengl=eufm10  scaled \magstep1
\font\sevengl=eufm7 scaled \magstep1
\font\fivegl=eufm5 scaled \magstep1

\newfam\dlfam  
\textfont\dlfam=\tendl \scriptfont\dlfam=\sevendl
\scriptscriptfont\dlfam=\fivedl
\newfam\glfam  
\textfont\glfam=\tengl \scriptfont\glfam=\sevengl
\scriptscriptfont\glfam=\fivegl

\def\draftdate{\number\month/\number\day/\number\year\ \ \ \hourmin }

\global\def\draftcontrol{0}
\catcode`\@=12
\def\tilde{\widetilde}
\def\hat{\widehat}
\documentstyle[12pt]{article}

\def\theequation{{\thesection.\arabic{equation}}}

\newcommand{\be}{\begin{eqnarray}}
\newcommand{\en}{\end{eqnarray}\vs 0.5 cm}

\newcommand{\no}{\noindent}
\newcommand{\vs}{\vskip}
\newcommand{\hs}{\hspace}

\newcommand{\NR}{{{\bf R}}}
\newcommand{\NZ}{{{\bf Z}}}

\newcommand{\NX}{{{\bf X}}}
\newcommand{\NY}{{{\bf Y}}}
\newcommand{\Na}{{{\bf a}}}
\newcommand{\Nx}{{{\bf x}}}
\newcommand{\Ny}{{{\bf y}}}
\newcommand{\Nv}{{{\bf v}}}

\newcommand{\Nf}{{{\bf f}}}

\newcommand{\Nz}{{{\bf z}}}
\newcommand{\Nk}{{{\bf k}}}

\newcommand{\qq}{\begin{eqnarray}}

\newcommand{\da}{\partial}
\newcommand{\ee}{{\rm e}}

\newcommand{\qqq}{\end{eqnarray}}

\newcommand{\tr}{\hbox{tr}}

\newcommand{\CA}{{\cal A}}
\newcommand{\CB}{{\cal B}}
\newcommand{\CC}{{\cal C}}
\newcommand{\CD}{{\cal D}}

\newcommand{\CF}{{\cal F}}

\newcommand{\CK}{{\cal K}}
\newcommand{\CL}{{\cal L}}
\newcommand{\CM}{{\cal M}}
\newcommand{\CN}{{\cal N}}
\newcommand{\CO}{{\cal O}}
\newcommand{\CP}{{\cal P}}

\newcommand{\CS}{{\cal S}}
\newcommand{\CT}{{\cal T}}

\newcommand{\s}{\hspace{0.05cm}}
\newcommand{\m}{\hspace{0.025cm}}

\newcommand{\hf}{{_1\over^2}}

\begin{document}

\begin{center}
\

\vs 1cm
{\Large{\bf{UNIVERSALITY in TURBULENCE\m:
\vs 0.2cm
an EXACTLY
SOLUBLE MODEL}}}\footnote{Lectures read by the first
author at the $34^{\m\rm th}$
School of Theoretical Physics in Schladming, Austria,
March 4-11, 1995}
\vs 1cm
{\large{Krzysztof Gaw\c{e}dzki}}
\vs 0.2cm
I.H.E.S., C.N.R.S.,
F-91440  Bures-sur-Yvette, France
\vs 0.5cm
{\large{Antti Kupiainen}}\footnote{Partially supported by
NSF grant DMS-9205296 and EC grant CHRX-CT93-0411}
\vs 0.2cm
Mathematics Department, Helsinki University,

PO Box 4, 00014 Helsinki, Finland
\end{center}
\date{ }

\vskip 1.3 cm

\begin{abstract}
\vskip 0.3cm

\noindent The present note contains the text of
lectures discussing the problem of universality in fully
developed turbulence. After a brief description
of Kolmogorov's 1941 scaling theory of turbulence
and a comparison between  the statistical approach
to turbulence and field theory, we discuss a simple
model of turbulent advection which is exactly
soluble but whose exact solution is still difficult
to analyze. The model exhibits a restricted
universality. Its correlation functions
contain terms with universal but anomalous scaling but with
non-universal amplitudes
typically diverging with the growing size of the system.
Strict universality applies only after such terms
have been removed leaving renormalized correlators
with normal scaling. We expect that the necessity of such
an infrared renormalization is a characteristic feature
of universality in turbulence.
\end{abstract}
\vs 1.6cm

\nsection{Introduction}
\vskip 0.5cm

One of the basic and most successful ideas in theoretical
physics has been that of universality. It states that
many systems with large or infinite number of degrees
of freedom in certain asymptotic regimes display similar
behaviors falling into general types. Such a situation
has been encountered
\vskip 0.3cm
\parbox{14.8cm}{in statistical mechanics of $2^{\rm nd}$
order phase transitions where the universality applies to
the long distance asymptotics of correlation functions
characterized e.\s g. by scaling exponents, independent
of numerous microscopic details of the systems \cite{DoGr6},}
\vskip 0.2cm
\parbox{14.9cm}{in field theory where the universality
implies the cutoff independence of the effective low energy
description \cite{DoGr6},}
\vskip 0.2cm
\parbox{14.9cm}{in the theory of $1^{\rm st}$ order phase
transitions exhibiting universal shapes of critical
droplets \cite{DShK},}
\vskip 0.2cm
\parbox{14.9cm}{in many-body condensed matter where
universality limits possible behaviors of
superconductors \cite{FMRT} or Hall fluids \cite{Fr},}
\vskip 0.2cm
\parbox{14.9cm}{in dynamical systems where it describes
types of behavior of families of maps under iteration \cite{Feig},}
\vskip 0.2cm
\parbox{14.9cm}{in non-linear PDE's where it determines
space and/or time asymptotics of solutions \cite{BrKup}.}
\vskip 0.3cm
\no The basic idea behind universality of certain behaviors
of systems with many degrees of freedom is that such behaviors
are governed by few relevant degrees of freedom with simple
dynamics. Those may be often exhibited using renormalization
group transformations wiping out the irrelevant details
of the system \cite{WilKo}\cite{Wil}.
\vskip 0.4cm

The question addressed in the present lectures is to what
extent such a picture applies to the fully developed
hydrodynamic turbulence.
It should be mentioned at the very beginning that the problem
is still wide open although the first theory arguing for
universality in turbulence, due to Kolmogorov \cite{K41},
dates from 1941. In the $1^{\m\rm st}$ lecture, we shall sketch
how the main claims of the Kolmogorov theory
relate to the cascade picture of energy
transfer in a turbulent flow. In the $2^{\m\rm nd}$
lecture, we shall briefly present the functional-integral approach
to the statistical theory of turbulence
governed by the Navier-Stokes equation trying to stress
the similarities and the differences with field theory.
In the $3^{\m\rm rd}$ and the $4^{\m\rm th}$ lecture,
we shall discuss a simple model
of turbulent advection of a scalar quantity,
known under the name
of passive scalar (PS). The model is a good playground for
testing the idea of universality in turbulence
since one may obtain closed expressions for the
stochastic initial data evolution and for the steady
state correlation functions of the scalar.
In the absence of external sources, the PS
with deterministic initial data undergoes
a non-universal diffusion which
averaged over initial data distributed in
a homogeneous way becomes a universal super-diffusion.
In the presence of random sources,
the steady state $2$-point correlator
exhibits an energy cascade
of the type expected in the Navier-Stokes turbulence.
The exact solution for the higher point
correlators are more difficult to analyze.
We examine the origin of
possible violations of strict universality of these
correlators. What emerges is a picture of restricted
universality applying to infrared renormalized correlators
from which few universal terms multiplied by
non-universal coefficients where removed.
In Conclusions, we summarize our discussion
and point to possible directions of the future
research.
\vs 0.3cm

The PS model,
although introduced long time ago \cite{Obu},
has become recently a subject of intensive study,
see \cite{Kr94}\cite{LPF} and also \cite{Majda}
(the latter, for a mathematical work on a simplified version
of the model).
In the final stages of the work on these notes,
we have received
preprints \cite{Russ}\cite{Proc} and \cite{Kra}.
The analysis of the first two has
large overlaps with ours although we disagree
with some conclusions of \cite{Russ} and differ
in some interpretations from \cite{Proc}.
\vs 0.3cm

We would like to thank the Mittag-Leffler Institute,
where our work on the passive scalar
model was started, for hospitality.
Discussions with Uriel Frisch,
Robbert Kraichnan, Itamar Procaccia and Achim Wirth
are kindly acknowledged.
\vs 1cm

\nsection{Energy cascade and Kolmogorov scaling}
\vs 0.5cm

The evolution of the local velocity field
\s$\Nv(t,\Nx)\s$ of the incompressible
fluid at points \s$\Nx\s$
of the $3$-dimensional space is described by
the Navier-Stokes (NS) equation
\qq
\da_t\Nv+(\Nv\cdot\nabla)\m\Nv
-\nu\m\Delta\Nv\s=\s-\nabla p+ \Nf
\label{NS}
\qqq
supplemented with the incompressibility condition
\qq
\nabla\cdot\Nv\s=\s0\ .
\label{InC}
\qqq
The presence of the dissipative term multiplied
by the viscosity $\nu$ of the fluid (with the dimension
${length}^2\over{time}$) distinguishes
the NS equation from the
Euler equation. $\Nf(t,\Nx)$ denotes the external force.
$p(t,\Nx)$ is the pressure (divided by the constant density)
and may be eliminated from both equations:
\qq
p(t,\Nx)\s=\s{_1\over^{4\pi}}\int{_{\nabla\cdot[\m\Nf(t,\Ny)-
(\Nv(t,\Ny)\cdot\nabla)\Nv(t,\Ny)\m]}
\over^{\vert\Nx-\Ny\vert}}\m
d^3\Ny\ .
\qqq
If the flow takes place in a finite volume, boundary conditions
should be specified. The size of the volume introduces an
"integral scale" $L$ into the problem and we shall assume that
the external force \s$\Nf\s$ acts only at length scales
comparable with $L$. Alternatively, we may consider flows
in infinite space excited by external force with Fourier
components concentrated in wavenumbers $\Nk$ with
$k\equiv\vert\Nk\vert\leq L^{-1}$. Under the rescaling
\qq
\tilde\Nv(t,\Nx)&=&\sigma\s\Nv(\tau t,s\Nx)\ ,\cr
\tilde\Nf(t,\Nx)&=&\sigma\tau\s\Nf(\tau t,s\Nx)\ ,\label{sc}\\
\tilde p(t,\Nx)&=&\sigma\tau s^{-1}p(\tau t,s\Nx)
\nonumber
\qqq
with \s$\sigma s/\tau=1\s$,
\s the viscosity \s$\nu\s$ is replaced
in Eq.\s\s(\ref{NS}) by \s$\tilde\nu=\tau s^{-2}\nu\s$.
\s It is then convenient to
introduce a dimensionless quantity, the Reynolds number
\qq
R\s=\s{_{V\s L}\over^{\nu}}\ ,
\qqq
where \s$V\s$ stands for the typical value of the velocity
differences over the integral scale \s$L$\m.
\s$R\s$ is invariant under the rescalings (\ref{sc}).
\vskip 0.4cm

The basic phenomenological observation is that the hydrodynamic
flows have very different behavior for small values of \s$R\s$
(of order $1$, say) where flows are regular (laminar) and
large values of \s$R\s$ (of order $10^3$ and more) where
the flows are very chaotic (turbulent), with complicated
phenomena occurring at intermediate values of \s$R\s$.
\s Here, we shall be interested in the regime of fully
developed turbulence with very large Reynolds numbers, ideally
in the limit \s$R\to\infty\s$.
\s It is sensible to use a statistical description of
the flows in this regime. The stochasticity may be introduced
into the description by considering random external force
\s$\Nf\s$ or studying the evolution of random initial data
or both. In a statistical description, the basic objects to
look at will be the velocity correlation functions given
by the expectation values of products of components of
velocities. We shall mostly look at equal time correlators
\qq
<\prod\limits_{n=1}^N v^{i_n}(t,\Nx_n)>
\label{etc}
\qqq
which, in the stationary state of turbulence sustained
by steady external forces, should be time independent.
Besides, we may expect that, far from the fluid boundaries,
the correlators (\ref{etc}) are translationally
and rotationally invariant, the property that
expresses the homogeneity and isotropy of the fully
developed turbulence. The most commonly studied
velocity correlators are the so called structure
functions
\qq
S_N(x)\s=\s<(\m\Nv(\Nx)-\Nv(0)\m)^N>\ .
\qqq
\vskip 0.3cm

One of the simplest consequences of the NS equation (\ref{NS})
may be obtained by taking its scalar product with \s$\Nv\s$
and integrating the result over the space. These results in
the relation
\qq
{_d\over^{dt}}\s\hf\smallint\Nv^2\s
=\s-{_\nu\over^2}\smallint(\nabla\Nv)^2\s
+\s\smallint\Nv\cdot\Nf
\label{EB}
\qqq
which expresses the energy balance: the time derivative
of energy on the right hand side is equal to the
difference of the injection rate \s$\int\Nf\cdot\Nv\s$
and the dissipation rate \s${_\nu\over^2}\int(\nabla\Nv)^2\s$
of energy. Taking averages in the stationary state,
we obtain
\qq
<\Nv\cdot\Nf>\s\s=\s\s
<{_\nu\over^2}\m(\nabla\Nv)^2>
\s\s\equiv\s\s\epsilon\ ,
\label{MEB}
\qqq
i.\s e. the equality of the (intensive) mean injection and
the mean dissipation rates of energy.
How is this energetic balance
realized over different modes of the fluid motion?
The energy injection takes place at the distances of
order of the integral scale by induction of big scale \s$L\s$
eddies. According to the picture of the turbulent
flow proposed in 1922 by Richardson \cite{Rich}, the big
eddies induce smaller eddies which, in turn, induce still
smaller eddies and so on transferring energy from large
to small distance scales. This process should not lead to a
loss of energy until sufficiently small
distance scales, say, smaller
than \s$\eta\s$, are reached.
On scales smaller than \s$\eta\s$,
\s the dissipative term \s$\nu\s\Delta\Nv\s$ of the NS
equation becomes important. One may formulate a more
quantitative version of this picture.
We shall follow here the discussion of the treatise \cite{Frisch}
which is an excellent reference to the problems
of well developed turbulence. The idea is to localize
the velocity field in the Fourier space.
For \s\s$\hat\Nv(\Nk)\equiv\int
\ee^{-i\m\Nk\cdot\Nx}\s\Nv(\Nx)\s
d^3\Nx\s\s$ define
\qq
\Nv_{_{\leq K}}(\Nx)\s=\s\smallint\limits_{\vert k\vert\leq K}
\ee^{i\m\Nk\cdot\Nx}\m\s\hat\Nv(\Nk)\s{_{d^3{\bf k}}
\over^{(2\pi)^3}}\ .
\qqq
A more detailed version of the energy balance equation (\ref{EB})
reads
\qq
{_d\over^{dt}}\s\hf\smallint\Nv_{_{\leq K}}^{\s\s2}\s
=\s-{_\nu\over^2}\smallint(\nabla\Nv_{_{\leq K}})^2\s
+\s\smallint\Nv_{_{\leq K}}\cdot\s\Nf\s-\s\smallint\Pi_{_K}\ ,
\label{DEB}
\qqq
where \s$\Pi_{_K}\s$ is the density of energy flux out
of the wave numbers with \s$\vert\Nk\vert\leq K\s$.
\s An explicit expression for \s$\Pi_{_K}\s$ may be obtained
from the NS equation (\ref{NS}) and may be argued to involve
mainly \s$\hat\Nv(\Nk)\s$ with \s$\vert\Nk\vert\cong K\s$,
\s i.\s e. to be aproximately local in the wavenumber space.
Taking the averages in the stationary state, we obtain the
relation
\qq
<\Nv_{_{\leq K}}\cdot\m\Nf>
\s\s=\s\s<{_\nu\over^2}\smallint
(\nabla\Nv_{_{\leq K}})^2>
\s+\s<\Pi_{_K}>
\label{MDEB}
\qqq
which states that the mean injection rate of energy into
the modes with \s$\vert\Nk\vert\leq K\s$ is equal to the
mean rate of
dissipation of energy \s$\epsilon_{_{\leq K}}\s\equiv
\s<{\nu\over2}\int(\nabla\Nv_{_{\leq K}})^2>\s$
in those modes plus the mean
flux of energy out of modes with \s$\vert\Nk\vert\leq K\s$.
According to the cascade picture of the energy transfer,
the dissipation rate \s$\epsilon_{_K}\s$ is
essentially zero for \s$K\ll \eta^{-1}\s$. \s Since the
injection takes place only around the integral scale,
the change of the mean injection rate
\s$\varphi_{_{\leq K}}\equiv\s\s
<\Nv_{_{\leq K}}\cdot\m\Nf>\s$ should be
negligeable for \s$K\gg L^{-1}\s$ so that \s$\varphi_{_{\leq K}}\s$
should be equal to its \s$K=\infty\s$ value \s$\epsilon\s$
for such \s$K\s$. \s It follows
from Eq.\s\s(\ref{MDEB}) that in the so
called {\bf inertial range} of
scales \s$L^{-1}\ll K\ll \eta^{-1}\s$, the mean energy flux in
the wavenumber space \s$\pi_K\equiv\s\s<\Pi_{_K}>\s$ is constant.
This is an important quantitative consequence of the
cascade picture of developed turbulence which, however,
is not easily verifiable directly.
\vs 0.4cm

Kolmogorov went further, postulating that in the inertial range
\s$\eta\ll x\ll L\s$, \s the velocity structure
functions may be expressed solely
as universal functions of the mean dissipation rate \s$\epsilon\s$
and the distance \s$x\s$, due to the locality
and self-similarity of the cascade. The shape of the structure
functions is then dictated
by the dimensional considerations. One obtains
\qq
S_N(x)\s=\s C_N\s\m\epsilon^{N/3}\s x^{N/3}
\label{SSF}
\qqq
with universal coefficients \s$C_N\s$ (the right hand side is the
only function of \s$\epsilon\s$ and \s$x\s$ with dimension
\s$({length\over time})^N\s$)\m.
\s Indeed, this behavior is consistent with the cascade picture.
With the typical velocity of size \s$l\s$ eddies
\s$v_l\sim S_N(l)^{1/N}\sim\epsilon^{1/3}l^{1/3}\s$,
\s their typical energy density \s$\hf\m v_l^2
\sim\epsilon^{2/3}l^{2/3}\s$
and their typical turnover time \s$t_l\sim l/v_l\sim\epsilon^{-1/3}
l^{2/3}\s$, one may estimate the energy flux from scales
longer than \s$l\s$ to those shorter than \s$l\s$
as \s$\sim v_l^2/t_l\sim\epsilon\s$ with no scale dependence.
\vskip 0.3cm

The scaling (\ref{SSF}) of the $2^{\m\rm nd}$ structure function
translates under the Fourier transform to the relation
\qq
{_1\over^{4\pi\m k^2}}\s E(k)\s\equiv\s
<\Nv(\Nk)\cdot\m\Nv(-\Nk)>\s\s=\s
\smallint\ee^{-i\m\Nk\cdot\Nx}\s\m S_2(x)\s\m d^3\Nx
\ \sim\ \epsilon^{2/3} k^{-11/3}
\qqq
so that the Kolmogorov theory prediction for
the energy spectrum \s$E(k)\s$ is
\qq
E(k)\ \sim\ \epsilon^{2/3}\s\m k^{-5/3}\ .
\label{ESP}
\qqq
The theory allows also to locate the scale \s$\eta\s$ at which
the dissipative effects become important, called usually
the Kolmogorov scale. Estimating the dissipation rate
on scale \s$l\s$ as \s$\sim \nu\m
({\displaystyle{v_l\over l}})^{^2}\s$,
\s \s$\eta\s$ should correspond to the scale at which
\s$\nu\m({\displaystyle{v_\eta\over\eta}})^{^2}\cong\epsilon\s$.
\s Using the relations
\qq
v_\eta\s\sim\s\epsilon^{1/3}\eta^{1/3}\ ,\hspace{1cm}
v_L\s\sim\s\epsilon^{1/3}L^{1/3}\ ,
\qqq
we infer that
\qq
\eta\s\sim\s({_{\nu}\over^{v_L L}})^{3/4}L\s=\s R^{-3/4}L\ .
\qqq
As should be expected, the Kolmogorov scale
decreases with growing Reynolds numbers,
i.e. with decreasing viscosity. In the limit \s$R\to\infty\s$,
the inertial range should invade all short scales. It does
not mean, however, that the behavior in that range may
be described by the Euler equation: the presence of even
very small viscosity is essential for the existence of
the stationary state of the turbulence by providing a
mechanism for removal from the system of the energy injected
by external forces.
\vs 0.4cm

The verifications of the
Kolmogorov theory is made difficult by the fact that
both in experiments in the atmosphere,
in wind tunnels or in liquid flows and in numerical
simulations it is hard to
obtain inertial ranges extending over many orders
of magnitude necessary to extract the characteristic
exponents of the structure functions. Nevertheless,
it seems that the behavior of low structure functions
well agrees with Kolmogorov's predictions
whereas for higher structure functions with $N\sim 10\s$
one observes values of exponents lower than predicted
(by $\sim 10\%$ or more). A multitude of more or less
{\it ad hoc} mechanisms has been proposed to explain the
possible departures from the Kolmogorov scaling for
the higher structure functions. The underlying idea
is that of intermittency limiting the turbulent
activity to a subset of temporal and spatial modes.
We shall not discuss the intermittent models
here referring the interested reader to \cite{Frisch}.
\vs 1cm

\nsection{Functional methods in turbulence}

The Kolmogorov theory has a phenomenological character.
It is strongly related to the cascade picture of turbulence
which was postulated essentially independently of the
NS equations. It may be compared to the
mean-field approach to critical phenomena
which leads to correct results
only in limited situations and which, in general, has to
be replaced by a more sophisticated
theory taking into account the
role of statistical fluctuations. It is then essential
to try to build a theory of fully developed turbulence
starting from the basic equations.
\vs 0.3cm

For any functional evolution equation of the type
\qq
\da_t \Phi\s=\s-\CF(\Phi)\s+\s F\ ,
\label{EV}
\qqq
where \s$F\s$ is a functional Gaussian process,
a Martin-Siggia-Rose
(MSR) formalism \cite{MSR} permits to obtain
formal functional integral expressions for the expectation
values of functionals \s$\CA(\Phi)\s$ in the stationary state.
This is done as follows. We rewrite
\qq
<\CA(\Phi)>\s\s=\m{_1\over \CN}\hs{-0.05cm}
\int\hs{-0.1cm}\CA(\Phi)\ \delta(
\da_t \Phi+\CF(\Phi)- F)\s\s
\det(\da_t+{_{\delta\CF(\Phi)}\over^{\delta\Phi}})
\ \ee^{\m-{1\over 2}\m(F\m,\s\CC^{-1} F)}\s\s
D\Phi\s\m DF,\hs{0.5cm}
\qqq
where \s$\CC\s$ is the covariance operator of the Gaussian
process \s$F\s$ and \s$\CN\s$ is the normalizing factor
given by a similar integral with \s$\CA=1\s$.
\s The role of the \s$\Phi\s$ integral
is to express \s$\Phi\s$ as a solution of the evolution
equation (\ref{EV}) (with a fixed initial condition).
If the functional \s$\CF(\Phi)\s$ is local in time
then, formally,
\qq
\det(\da_t+{_{\delta\CF(\Phi)}\over^{\delta\Phi}})\s=\s
\det(\da_t)\s\s\det(1+\da_t^{\s-1}
{_{\delta\CF(\Phi)}\over^{\delta\Phi}})
\s=\s\det(\da_t)\s\s
\ee^{\s\hf\s{\bf tr}\s\m{\delta\CF(\Phi)/\delta\Phi}}
\qqq
since \s$\da_t^{-1}(t,t')=\theta(t-t')\s$ and
\s$\delta\CF(\Phi)/\delta\Phi\s$ is proportional to
\s$\delta(t-t')\s$.
\s Alternatively, one may use the (Berezin) functional
integral over anticommuting fields to exponentiate the
determinant. Rewriting also the delta function
\s$\delta(\da_t \Phi+\CF(\Phi)\s-\s F)\s$ as an oscillatory
integral, we obtain
\qq
<\CA(\Phi)>\s\s=\s{_1\over \CN}
\m\int\CA(\Phi)\ \ee^{\s i\s(A\s,\m\s
\da_t \Phi+\CF(\Phi)\s-\s F)\s+\s{1\over 2}\m
{\bf tr}\s\m{\delta\CF(\Phi)/\delta\Phi}\s
-\s{1\over 2}\m(F\m,\s\CC^{-1}F)}\ DA\s\s D\Phi\s\s DF\ .\ \
\qqq
Finally, the Gaussian integral over \s$F\s$ leads to the
relation
\qq
<\CA(\Phi)>\s\s=\s{_1\over \CN}
\m\int\CA(\Phi)\ \ee^{\s i\s(A\s,\m\s
\da_t \Phi+\CF(\Phi))\s+\s{1\over 2}\m
{\bf tr}\s\m{\delta\CF(\Phi)/\delta\Phi}\s
-\s{1\over 2}\m(A\m,\s\CC\m A)}\ DA\s\s D\Phi\ .
\label{MSR}
\qqq
\vs 0.3cm

The MSR formalism applies to the Langevin equation
describing the approach to equilibrium in statistical
mechanics and Euclidean field theory. In this case,
the functional \s$\CF(\Phi)\s$ appearing in the
evolution equation (\ref{EV}) is of the gradient type:
\qq
\CF(\Phi)={_{\delta S(\Phi)}\over^{\delta\Phi}}
\qqq
and the Gaussian process \s$F\s$ is the  white noise, i.\s e.
its covariance \s$\CC\s$ is the identity operator.
For example, for the scalar \s$\phi^4\s$ theory,
\qq
S(\Phi)\s=\s\int\left(\hf\m \nabla\Phi)^2+\hf\m m^2\Phi^2+
\lambda\m\Phi^4\right)\s dt\s\s d^d\Nx\ .
\qqq
Perturbative treatment of the functional integral (\ref{MSR})
based on expanding the exponential of the terms of order
higher then $2$ in \s$\Phi\s$ and \s$A\s$ into a power
series gives rise to the perturbative expansion for
the correlation functions.
The latter reduces for the equal-time
correlators to the perturbative expansion for
the equilibrium expectations in the Gibbs measure
\s\s${1\over\CN}\s\s\ee^{-S(\phi)}\s\s D\phi\s\s$ with
\s$\phi\s$ as \s$\Phi\s$ but without the time dependence.
\vs 0.3cm

One may rewrite the NS equations (\ref{NS}) and (\ref{InC})
as a single evolution equation
\qq
\da_t\Nv\s=\s-P\s(\Nv\cdot\nabla)\m\Nv\s
+\s\nu\s\Delta\Nv\s+\s P\m\Nf
\label{EV1}
\qqq
in the space of divergence-free vector fields, where \s$P\s$
stands for the orthogonal projection on such fields.
Taking the external force \s$\Nf\s$ random Gaussian,
we end up in the setup formally resembling the
Langevin dynamics.
The important differences between the latter
and Eq.\s\s(\ref{EV1}) are, however, that, first, the term
\s$P\s(\Nv\cdot\nabla)\m\Nv\s$ is not of the gradient type
and, second, that we want to look at the noise \s$\Nf\s$
with the Fourier components concentrated in low wavenumbers,
i.\s e. with the kernel of the covariance in the position space
close to a constant as opposed to the
delta function kernel for the Langevin equation. As a result,
the role of the noise in the NS equation is to inject
at small wavenumbers the energy which is then transferred
to larger wavenumbers by an essentially deterministic process
leading to the non-zero energy flux in the stationary state
whereas in the Langevin dynamics, the noise plays an essential
role on all scales resulting in the thermal state with no
non-zero energy fluxes between scales.
\vskip 0.3cm

It is sometimes pointed \cite{Polya} that one encounters
non-vanishing momentum-space fluxes also in the "equilibrium"
field theory in the presence of anomalies. For the sake
of illustration, let us consider the chiral anomaly in two
space-time dimensions. The chiral charge density
of the Dirac field is
\qq
J^{50}(t,x)\m=\m\bar\psi(t,x)\m\gamma^5\gamma^0\m\psi(t,x)\ .
\qqq
The anomaly calculation shows that in the external
abelian gauge field in two space-time dimensions
\qq
{_d\over^{dt}}\s<\smallint J^{50}(t,x)\s
dx>_{_{\leq K}}
\s\s=\s{_1\over^{2\pi}}\smallint\epsilon^{\mu\nu}\s
F_{\mu\nu}(t,x)\s\m dx\ ,
\qqq
and the right hand side is \s$K\s$ independent.
Here the subscript \s$\leq K\s$ denotes a gauge-invariant
(e.g. Pauli-Villars) subtraction at momentum-scale \s$K\s$.
\s In other words, the flux
of the  chiral charge out of the modes with momenta \s$\leq K\s$
is constant in \s$K\s$. \s It is not clear what lessons
may be drawn from this analogy for understanding
the inertial range of the fully developed turbulence.
\vskip 0.3cm

Applying the MSR formalism to Eq.\s\s(\ref{EV1}), we obtain
for the velocity correlation functions the expression
\qq
<\prod\limits_{n=1}^N v^{i_n}(t_n,\Nx_n)>\s
=\m{_1\over \CN}
\m\int\prod\limits_{n=1}^N v^{i_n}(t_n,\Nx_n)\
\ee^{\s i\s(\m\Na\m,\m\m
\da_t \Nv+(\Nv\cdot\nabla)\m\Nv\m-\m\nu\s\Delta\Nv\m)\m
-\m{1\over 2}\m(\Na\m,\s\CC\m \Na)}\ D\Na\s\s D\Nv\s,\hs{0.5cm}
\label{NSMSR}
\qqq
where the \s$\Na$- and \s$\Nv$-integrals are over divergence-free
vector fields \s(\s$\m\tr\s\m{\delta\m(\Nv\cdot\nabla)\m\Nv\over
\delta\Nv}\s$ drops out as it vanishes
because of symmetry reasons).
Separating the expression under the
exponential into the quadratic part
\s\s$i\s(\m\Na\m,\m\m
\da_t \Nv\m-\m\nu\s\Delta\Nv\m)\m
-\m{1\over 2}\m(\Na\m,\s\CC\m \Na)\s\s$ and the cubic
one \s\s$i\s(\m\Na\s,\m\m(\Nv\cdot\nabla)\m\Nv\m)\s\s$
and expanding the exponential of the latter into the
power series, we may generate the perturbation expansion
for the velocity correlators, just like for the
correlation functions of the Langevin evolution.
The terms of the expansion may be represented by
Feynman diagrams with the propagators read off the
quadratic part under the exponential and the vertices
off the cubic part. A closer look into the perturbation
expansion generated this way shows that it is plagued
by infrared and ultraviolet divergences (the latter in
the limit of vanishing \s$\nu\s$)\m.
\s Various resummation techniques, mostly inspired
by approaches used in field theory, were tried with
the aim to improve the convergence, without
complete success. Let us mention here another approach,
originally labeled quasi-Lagrangian,
which seems more promising. This approach has been developed
over years by the Novosibirsk and more recently
the Weizmann Institute schools
\cite{BerlLv}\cite{Lv}\cite{LvProc}.
\vskip 0.4cm

The idea consists of describing the turbulent flow in
the frame moving with one of its points. Suppose that,
for given velocity field \s$\Nv(t,\Nx)\s$, \s$\Nx(t)\s$
is the solution of the ODE
\qq
{_{d{\bf x}}\over^{dt}}\m=\m\Nv(t,\Nx)
\qqq
with the initial condition \s$\Nx(t_0)=\Nx_0\s$.
\s$\Nx(t)\s$ is the trajectory of the material point of
the fluid located at \s$\Nx_0\s$ at time \s$t_0\s$.
\s The velocity field, force and pressure in the frame moving
with the fluid point are
\qq
\Nv'(t,\Nx)&=&\Nv(t,\Nx+\Nx(t))\ ,\label{qlv}\\
\Nf'(t,\Nx)&=&\Nf(t,\Nx+\Nx(t))\ ,\label{qlf}\\
p'(t,\Nx)&=&p(t,\Nx+\Nx(t))\ ,
\qqq
and they satisfy the equations following
from the NS equations for \s$\Nv\s$:
\qq
&\da_t\Nv'+((\Nv'-\Nv'_0)\cdot\nabla)\m\Nv'
-\nu\m\Delta\Nv'\s=\s-\nabla p'+ \Nf'\ ,
\label{qNS}\\
&\nabla\cdot\Nv'\s=\s0\ ,
\label{qInC}
\qqq
where \s$\Nv'_0(t)\equiv\Nv'_0(t,0)\s$.
\s Eqs.\s\s(\ref{qNS}) and (\ref{qInC})
may be again written as a single
evolution equation in the space of divergence-free
vector fields:
\qq
\da_t\Nv'\s=\s-P\s((\Nv'-\Nv'_0)\cdot\nabla)\m\Nv'\s
+\s\nu\s\Delta\Nv'\s+\s P\m\Nf'
\label{qEV1}
\qqq
to which the MSR formalism may be applied if we assume
that \s$\Nf'\s$ is a Gaussian random field. This
is not the same as a similar assumption about \s$\Nf\s$
but as long as the Fourier components of the force
are concentrated in small wavenumbers, the
details of the force distribution should not matter
in the inertial range. In such a situation, one may
expect that the equal-time velocity correlators
in the stationary state of the evolution equation
(\ref{qEV1}) coincide in the inertial range with
those corresponding to the equation (\ref{EV1}).
The perturbation expansion generated by the MSR technique
for the \s$\Nv'$-correlators has the same propagators
but modified vertices which break the translation
invariance and make this way the quasi-Lagrangian
expansion more difficult to analyze. Nevertheless
it seems that the quasi-Lagrangian approach
leads to drastic improvement in convergence
of the (Schwinger-Dyson resummation) of the
perturbative expansion \cite{BerlLv}\cite{LvProc}
and it is possible that it provides an important
clue to the understanding of the
statistical properties of the turbulent flow.
\vs 1cm

\nsection{Passive scalar}
\subsection{Definition of the model}
\vs 0.5cm

In view of the difficulties encountered by the theory
of the Navier-Stokes turbulence, it may be useful
to search for simpler systems exhibiting some of
the aspects of full-fledged turbulence but easier
to control. Such a system seems to be provided by
a model of passive advection in a random velocity
field \s$\Nv(t,\Nx)\s$ of a scalar quantity \s$T\s$
whose density \s$T(t,\Nx)\s$ satisfies the equation
\qq
\da_t\m T\s+\s(\Nv\cdot\nabla)\m T\s-\s\nu\Delta T\s=\s f\ ,
\label{PS}
\qqq
where now \s$\nu\s$ denotes the molecular diffusivity
of the scalar \s$T\s$
and \s$f(t,\Nx)\s$ describes the external sources.
In the ideal situation, we should consider \s$\Nv\s$
as the velocity of the NS turbulent flow,
but we shall, instead, assume that \s$\Nv(t,\Nx)\s$
is a centered Gaussian field with the covariance
\qq
< v^i(t,\Nx)\m\s v^j(t',\Nx')>
\s=\s\delta(t-t')\s\s D^{ij}(\Nx-\Nx')\ ,
\label{42}
\qqq
i.\s e. white noise in time. The spatial part of the
covariance will be taken, for concreteness, as
\qq
D^{ij}(\Nx)\s=\s D_0\m\int\ee^{\m i\m\Nk\cdot\Nx}
\s\s(\Nk^2+m^2)^{-(3+\kappa)/2}\s\s(\delta^{ij}-k^ik^j/\Nk^2)
\s\s{_{d^3{\bf k}}\over^{(2\pi)^3}}\ ,
\label{43}
\qqq
where the transverse projector in the Fourier space
ensures the incompressibility of \s$\Nv\s$.
Small $m^2$ should be viewed as an infrared cutoff
making the integral convergent for \s$0<\kappa<2\s$.
Note that the dimension of \s$D_0\s$ is \s${length^{2-\kappa}
\over time}\s$. \s For the $2^{\m\rm nd}$ velocity
structure function we obtain
\qq
<(\Nv(t,\Nx)-\Nv(t,0))^2>
\ \sim\ D^{ii}(0)-D^{ii}(\Nx)\s\equiv\s \tilde D^{ii}(\Nx)
\qqq
which, for \s$0<\kappa<2\s$, has the \s$m\to 0\s$
limit proportional to \s$\vert\Nx\vert^{\kappa}\s$.
i.\s e. growing with the distance. More exactly,
\qq
D^{ij}(0)\s=\s{_{16\s\m\Gamma(\kappa/2)}\over^{3\pi^{3/2}
\s\Gamma(\kappa/2)}}\s\m\delta^{ij}\s\m D_0\s\m m^{-\kappa}\ ,
\label{div}
\qqq
i.\s e. it diverges with \s$m\to 0\s$, but
\qq
\lim_{m\to 0}\s\s\tilde D^{ij}(\Nx)\s=\s
D_1\s\left((2+\kappa)\s\delta^{ij}\m
\vert\Nx\vert^{\kappa}\s-\s\kappa\s\m x^ix^j\m
\vert\Nx\vert^{\kappa-2}\right)
\label{td}
\qqq
and it is a homogeneous function of \s$\Nx\s$.
\s$D_1\s\equiv\s
{{\Gamma((2-\kappa)/2)}\over{2^{2+\kappa}\pi^{3/2}
\kappa(3+\kappa)}}\s\m D_0\s$.
\s Note the singularities at $0$ and $2$ in the
\s$\kappa$-dependence. The Kolmogorov scaling
(\ref{SSF}) corresponds to \s$\kappa={2\over3}\s$
which lies in the interval \s$0<\kappa<2\s$ that we
shall consider. It should be mentioned, however,
that the time-decorrelation,
evenness and Gaussian character of the velocity
distribution is a rather drastic departure from the
statistical properties of the turbulent NS velocities.
Nevertheless, the model, dating back to 1949 \cite{Obu},
possesses an interest of its own
and has been used to describe a variety of phenomena
like poluant or tracer transport and is closely related
to models describing forced
diffusion through porous media \cite{GlSh}.
\vs 0.4cm

We shall take the field \s$f(t,\Nx)\s$
describing the external sources also random Gaussian,
independent of \s$\Nv\s$,
with mean zero and covariance
\qq
< f(t,\Nx)\s\m f(t',\Nx')>
\s=\s\delta(t-t')\ \CC({_{\Nx-\Nx'}\over^L})
\m\equiv\s\delta(t-t')\ \CC_L(\Nx-\Nx')
\label{force}
\qqq
where \s$\CC\s$ is a real positive-definite function
from the Schwartz space \s$\CS(\NR^3)\s$
and \s$L\s$ is the integral scale.
Sometimes, it will be useful to consider also the advection
of a complex scalar in which case \s$\CC\s$ may be
a complex positive-definite function.
We would like to study the statistical properties
of the solutions of Eq.\s\s(\ref{PS}) in the regime
of small \s$\nu\s$, small \s$m\s$ (which may viewed as
the inverse of another integral scale) and large \s$L\s$
(in the limit \s$L\to\infty\s$, the field \s$f\s$ has
only the zero wavenumber component). In particular,
the universality question for the PS
may be formulated as
inquiring about the existence of the limit
of the correlation functions
\qq
<\prod\limits_{n=1}^N T(t_n,\Nx_n)>
\label{CFPS}
\qqq
in a stationary state of the system when
\s$\nu,\m m,\m L^{-1}\m\to 0\s$ and about the independence
of such a limit of the shape of the source covariance \s$\CC\s$.
\vs 0.7cm

\subsection{Exact solution}
\vskip 0.3cm

The big advantage of the PS model over the
NS case is that one may obtain closed expressions for
the correlation functions of the scalar
(the assumed time-decorrelation
of the velocity field plays here the crucial role).
For $2$-point function, this was first done
in \cite{Kraich}\footnote{We thank U. Frisch
for explaining to us the $4$-point function case}.
In this sense the model is exactly soluble.
In order to work out its exact solution,
let us rewrite the basic evolution equation
(\ref{PS}) of the scalar in a general form
\qq
\da_t\m T\s+\s\beta\m T\s+\s\nu\m \alpha\m T=\s f\ ,
\label{FDE}
\qqq
where \s$\beta\s$ is a skew-symmetric random operator
(\s$\beta\s=\s\Nv\cdot\nabla\s$ in our case) and
\s$\alpha\s$ is a positive operator (\s$\alpha=-\Delta\s$).
The \s$\beta\s$ term in the equation is conservative
whereas the \s$\alpha\s$ one is dissipative (for
the energy \s$\hf\m\|T\|^2\s$)\m.
\s For the sake of simplicity, we shall argue
in the finite-dimensional case, i.\s e. when
\s$T(t)\s$ and \s$f(t)\s$ take values
in \s$\NR^d\s$ and \s$\beta(t)\s$
and \s$\alpha\s$ are, respectively,
skew-symmetric and positive
symmetric \s$d\times d\s$ matrices. For the
covariances of independent centered
Gaussian processes \s$\beta(t)\s$
and \s$f(t)\s$, \s we shall write
\qq
<\beta_{ij}(t)\s\m\beta_{kl}(s)>&=&\delta(t-s)\s\s
\CB_{ij,kl}\ ,\cr\cr
< f_i(t)\s\m f_j(s)>\ &=&\delta(t-s)\s\s\CC_{ij}\ .
\label{Cov}
\qqq
In finite dimensions, the original stochastic PDE
becomes a stochastic ODE which is easier to handle.
Let us simplify the matter further by assuming,
for a moment, that
\s$\beta(t)\s$ and \s$f(t)\s$ are continuous
functions. Then the solution of Eq.\s\s(\ref{FDE}) with
the initial condition \s$T_0\s$ at \s$t=t_0\s$
takes the well known form
\qq
T(t)\s=\s R(t,t_0)\s\m T_0\s
\m+\s\int_{t_0}^t\hs{-0.15cm} R(t,s)\s
f(s)\s ds\ ,
\label{FDS}
\qqq
where \s$R(t,t_0)\s$ is given by
the ordered exponential (\s$t\geq t_0\s$)
\qq
R(t,t_0)\ =\ \CT\ \ee^{\m
-\int_{t_0}^t(\m\nu\m\alpha\m+\m\beta(\tau)\m)\s\m d\tau\s}
\qqq
($\CT\s$ stands for the time ordering).
\s$R(t,t_0)\s$ may be also expressed by the following
limiting formula\s:
\qq
R(t,t_0)\s\s=\s\s\lim_{Q\rightarrow\infty}\ \m
\exp[\m-\smallint_{I_{_Q}\s}
(\nu\m \alpha+\beta(\tau))\s d\tau\s]
\s\s\m\exp[\m-\smallint_{I_{_{Q\hs{-0.05cm}-1}}}
(\nu\m \alpha+\beta(\tau))\s d\tau\s]
\ \cdots\ \s\cr
\cdots\ \exp[\m-\smallint_{I_{_1}\s}(\nu\m \alpha+
\beta(\tau))\s d\tau\s]
\s\s\equiv\s\s\lim_{Q\rightarrow\infty}\ \m
R_Q(t,t_0)\ ,
\label{Trott}
\qqq
where \s$(I_q)$ is a time ordered
partition of the interval
\s$[t_0,t]\s$ into subintervals
of length \s$(t-t_0)/Q\s$.
\vs 0.3cm

In the random case, the white
noise stochastic processes \s$\beta(t)\s$ and
\s$f(t)\s$ do not have continuous realizations.
They need smearing to give genuine random variables.
It is easy to see that
\s$\int_{I}\beta(\tau)\s d\tau\s\equiv\s\beta^I\s$
\s where \s$I\s$ are intervals of time, are
already genuine Gaussian random variables which are
mutually independent
for non-intersecting or adjacent intervals:
\qq
<\beta^I_{ij}\s\s\beta^J_{kl}>\s=\s\vert I\cap J\vert
\ \CB_{ij,kl}\ .
\qqq
Let us denote by \s$\CB\s$ the covariance operator
acting on skew-symmetric matrices by
\qq
(\CB\s\beta)_{ij}\s=\s\sum_{ijkl}\CB_{ij,kl}\s\beta_{kl}\ .
\label{CB}
\qqq
Necessarily, \s$\CB\geq 0\s$.
We shall also need below a contracted version
\s$B\s$ of \s$\CB\s$
with the matrix elements
\qq
B_{ij}\s=\s-\m\hf\m\sum\limits_{k}\CB_{ik,kj}\ .
\label{B}
\qqq
$B\s$ is a symmetric operator acting in \s$\NR^d\s$.
\s Note that
\qq
<((\beta^I)^t\s\beta^I)_{ij}>\s\s=
-<(\beta^I\s\beta^I)_{ij}>\s\s
=\s2\m\vert I\vert\s B_{ij}
\label{B1}
\qqq
from which it follows that \s$B\geq 0\s$.
In the random case with white noise \s$\beta(t)\s$, \s
the approximate evolution operators
\s$R_Q(t,t_0)\s$ under the
limit on the right hand side
of Eq.\s\s(\ref{Trott}) still make sense for each \s$Q\s$
as random (matrix valued) variables and
one can show that
\vs 0.8cm

\no{\bf Proposition 1}.
\qq
\lim_{Q\rightarrow\infty}
< R_Q(t,t_0)>\ =\m\s \ee^{\m-\m(t-t_0)\m(\nu\m\alpha
+B)\m}\ \equiv\
\ \ee^{\m-\m(t-t_0)\m\CM_1\m}\ .
\label{Prop1}
\qqq
\vs 0.7cm
\no In order to indicate how the above result arises,
consider the perturbation expansion for \s$R(t,t_0)\s$
in powers of \s$\beta\s$
\qq
R(t,t_0)\s=\sum\limits_{m=0}^\infty(-1)^m\hs{-0.06cm}
\int\limits_{t_0}^td\tau_m\int\limits_{t_0}^{\tau_m}d\tau_{m-1}
\cdots\int\limits_{t_0}^{\tau_2}d\tau_1
\m\s\ee^{-(t-\tau_m)\s\nu\m\alpha}\m\beta(\tau_m)\m\s\ee^{
-(\tau_m-\tau_{m-1})\m\nu\m\alpha}\s\beta(\tau_{m-1})\cr
\cdots\ \beta(\tau_1)\s\m\ee^{-(\tau_1-t_0)
\m\nu\m\alpha}\ .\hs{0.9cm}
\label{pee}
\qqq
In the white noise expectation value of the right hand side
computed with the use of the Wick Theorem, only
neighboring \s$\beta(\tau_m)\s$ may be paired
forcing the respective times \s$\tau_m\s$ to be equal.
One obtains this way the perturbation expansion of
\s$\ee^{\m-\m(t-t_0)\m(\nu\m\alpha+B)}\s$ in powers of
\s$B\s$. Proposition 1 is a rigorous version of
this result, not very difficult to prove \cite{my}.
\vs 0.4cm

Similarly, one may calculate expectations
of products of matrix elements of \s$R_Q(t,t_0)\s$.
\s We shall use the tensor
product notation \s$R_Q(t,t_0)^{\otimes\m N}\s$
as a bookkeeping device for all such products.
Let us denote by
\s$\CK\s$ the symmetric operator on
\s$\NR^d\otimes\NR^d\cong
{\rm Mat}_{d\times d}\s$
built of the coefficients \s$\CB_{ij,kl}\s$
so that
\qq
(\m\CK\s g)_{ik}\s=\s\sum\limits_{jl}\CB_{ij,kl}\s g_{jl}\ .
\label{CK}
\qqq
Note the difference between \s$\CK\s$
and the covariance
operator \s$\CB\s$ of the process \s$\beta(t)\s$
defined in Eq.\s\s(\ref{CB}).
In what follows an important role will be played by
the \s$N$-body symmetric operators \s$\CM_N\s$ acting
in \s$(\NR^d)^{\otimes\m N}
\cong\NR^{Nd}\s$ defined by
\qq
\CM_N\ \s\equiv\s\ \sum\limits_{n=1}^N\s\s (\CM_1)_n \s
\s\m-\hs{-0.1cm}
\sum\limits_{1\leq n<n'\leq N}\hs{-0.3cm}(\m\CK)_{n,n'}
\ ,
\label{CM}
\qqq
where \s$\CM_1=\nu\m \alpha+B\s$,
\s$(\CM_1)_n\s$ denotes \s$\CM_1\s$
acting in the \s$n^{\m{\rm th}}\s$ factor of the tensor
product \s$(\NR^d)^{\otimes\m N}\s$ and, similarly,
\s$(\m\CK)_{n,n'}\s$ stands for \s$\CK\s$ acting in the
\s$n^{\m{\rm th}}\s$ and \s$n'^{\s\rm{th}}\s$ factors
of \s$(\NR^d)^{\otimes\m N}\s$                        .
\vs 0.8cm

\no{\bf Proposition 2}. \s For \s$N=2,3,\dots\s$,
\qq
\lim_{Q\rightarrow\infty}\s\s< R_Q(t,t_0)^{\otimes N}>
\ =\ \ee^{\m-\m(t-t_0)\s\CM_N}\ .
\label{Prop2}
\qqq
Again, this result is not surprising if we use
the perturbation expansion (\ref{pee}) for
each \s$R(t,t_0)\s$ and note that the white
noise expectation may pair now, besides
the nearest neighbor \s$\beta(\tau_m)\s$
in the perturbative terms of the same
\s$R(t,t_0)\s$ also \s$\beta(\tau_m)\s$
corresponding to different \s$R(t,t_0)\s$,
\s resulting in the perturbative expansion
of \s$\ee^{\m-\m(t-t_0)\m\CM_N}\s$ in
powers of operators \s$B\s$ and \s$\CK\s$.
In fact, one may show \cite{my}
that \s$R_Q(t,t_0)\s$ converge to a limit
in all \s$L^p\s$ spaces, \s$p<\infty\s$,
of the white noise \s$\beta\s$.
\s Naturally, we shall denote the limit by \s$R(t,t_0)\s$.
\s$R(t,t_0)\s$ are contracting operators,
they are continuous in \s$t,\m t_0\s$
in the \s$L^p$-norms and satisfy the
composition law. We shall take then \s$T(t)\s$
given by Eq.\s\s(\ref{FDS}) with
the limiting \s$R(t,t_0)\s$ as the
solution of the stochastic equation
(\ref{FDE}) with initial value \s$T_0\s$.
\vs 0.4cm

In the absence of the external sources
the time evolution of the equal-time correlators
of the scalar \s$T\s$
simply reduces to the relations
\qq
< T(t)^{\otimes\m N}>\ \s=\s\s\ee^{\m-\m(t-t_0)\m\CM_N}\s
T(t_0)^{\otimes\m N}\ ,
\label{decay}
\qqq
hence it is given by the
semigroups \s$(\ee^{\m-\m t\s\CM_N})_{t\geq0}\s$
which may be studied by looking at the
Fourier transforms of the semigroups
i.e.\s\s at the resolvents \s$(\m i\m\omega+\CM_N)^{-1}\s$.
In any case, the properties of the time evolution
of the statistics of initial data are determined by
the spectral properties
of the symmetric operators \s$\CM_N\s$.
\s It is easy to see that the operators
\s$\CM_N\s$ satisfy the inequality
\qq
\CM_N\ \geq\ \nu\m\sum\limits_{n=1}^N(\alpha)_n\ .
\label{pos}
\qqq
An important consequence of this inequality is
the exponential time decay of the correlation functions
\s$< T(t)^{\otimes\m N}>\s$ as long as the
dissipation is present
in the equation (\ref{FDE}) (recall that we are looking at
the finite-dimensional case). The decay rate is given by the
lowest eigenvalues \s$\lambda_{N,0}\s$ of \s$\CM_N\s$
and is bounded below by \s$N\m\nu\m\mu_0\s$ where
\s$\mu_0>0\s$ is the lowest eigenvalue of \s$\alpha\s$.
If we remove the dissipative term from the
equation by setting \s$\nu=0\s$ then the long time behavior of
\s$< T(t)^{\otimes\m N}>\s$ will have a persistent
tail since the operators \s$\CM_N\s$ develop
zero modes when \s$\nu\s$ is taken to zero.
Suppose now that the initial data \s$T(t_0)\s$
are also random  and independent of \s$\beta(t)\s$
for \s$t\geq t_0\s$.
Averaging Eq.\s\s(\ref{decay}) over \s$T_0\s$, \s we obtain
\qq
< T(t)^{\otimes\m N}>\ \s=\s\s\ee^{\m-\m(t-t_0)\m\CM_N}\s
< T(t_0)^{\otimes\m N}>\ .
\label{decay1}
\qqq
A natural problem is the existence
of an invariant measure
on the space of initial data.
It follows from relations (\ref{pos}) and (\ref{decay1})
that there are no invariant measures with finite moments
on the space
of initial data as long as \s$\nu>0\s$. On the other
hand, when \s$\nu=0\s$, \s i.\s e. in the absence
of dissipation, one may show that any measure
preserved by a.\s a. rotations \s$\ee^{\m\beta^I}\s$
is invariant and that this property characterizes the
invariant measures with finite characteristic functions
\cite {my}. In particular, the Gibbs
measures \s$\sim\s\ee^{\m-\m{\rm const}.
\m\|T\|^2}\s\m d^dT\s$ are invariant under
the flow (\ref{FDE}) when \s$\nu=0\s$.
As we shall see, the situation is quite different
for infinite number of degrees of freedom.
\vs 0.5cm

In the presence of white-noise sources and of dissipation,
the $2$-point equal time correlation function
evolves according to the equation
\qq
< T(t)^{\otimes\m{2}}>\s&=&\s\ee^{\m-\m(t-t_0)\s\CM_2}\s\s
T(t_0)^{\otimes\m 2}\s+\s\smallint_{t_0}^t ds\s\s
\ee^{\m-\m(t-s)\s\CM_2}\s\s\CC\ \cr
&=&\s\ee^{\m-\m(t-t_0)\s\CM_2}\s\s
T(t_0)^{\otimes\m 2}\s+\s(1-\ee^{\m-(t-t_0)\m\CM_2})
\s\CM_2^{-1}\s\m\CC
\label{2point}
\qqq
which is a solution of the differential equation
\qq
\da_t\s< T(t)^{\otimes\m{2}}>\s=\s-\CM_2\s
< T(t)^{\otimes\m{2}}>\s+\s\s\CC\ .
\label{2poin}
\qqq
When \s$t_0\rightarrow-\infty\s$, \s the term with
\s$T(t_0)\s$ disappears due to the positivity of
\s$\CM_2\s$ (for \s$\nu>0\s$) \m and we obtain
\qq
< T^{\otimes\m{2}}>\ =\ \CM_2^{\s-1}\s\CC\ .
\label{2pst}
\qqq
This is the time independent $2$-point function in a steady state
in which the quantities of the energy injected
by the force and dissipated balance each other.
Similarly, for the general
equal time correlators
in the presence of the external sources and
of dissipation, the dependence on the initial conditions
is wiped out when \s$t_0\to-\infty\s$ due to the positivity
of all \s$\CM_N\s$. \s The higher stationary state
equal-time correlation
functions take the form:
\qq
< T^{\otimes\m{2M+1}}>&=&\s0\ , \cr\cr
< T^{\otimes\m{2M}}>\ \ \s&=&
\sum\limits_{\sigma\s\in\s\CS_{2M}\s/\s\CS_M\times
\CS_2^M}\hs{-0.3cm}\sigma\ F_{2M}\ ,
\label{pairi}
\qqq
where \s$F_{2M}\m\in\m(\NR^d)^{\otimes\m{2M}}\s$
is a single $2$-particle channel contribution.
Permutations \s$\sigma\s$ interchange the factors
in \s$(\NR^d)^{\otimes\m{2M}}\s$,
\s$\CS_M\s$ is viewed as a set of permutations of the pairs
\s$(2m-1,2m)\s,\ m=1,\dots,M\s$, \s and
\s$\CS_2^M\s$ as that of permutations acting
within the pairs so that the sum is over the pairings
of \s$\{1,\dots,2M\}\s$ or $2$-particle channels.
\qq
F_{2M}\ =\ \CM_{2M}^{-1}\sum\limits_{\sigma\in\CS_M}
\hs{-0.1cm}\sigma\m\left(\m
(\m\CM_{2M-2}^{\ \m-1}
\otimes\m 1_2\m)\ \cdots\
(\m\CM_2^{\s-1}\otimes\m 1_{2M-2}\m)\m\right)\s
\CC^{\otimes\m M}\ .\hs{0.4cm}
\label{1chan}
\qqq
Exact expressions for non-equal times \s$N$-point
correlators are also easy
to obtain. They involve also the heat
operators \s$\ee^{-t\CM_{N'}}\s$
with \s$N'<N$.
\vs 0.4cm

Let us discuss briefly some multibody combinatorics
behind the solution (\ref{1chan}).
Let us denote by \s$\CP_{2M}\s$ the set of pairs
\s$(2m-1,2m)\s,\ m=1,\dots,M\s$.
\s We shall define
the connected parts \s$F^c_{2M}\s$ of the single channel
correlator by the inductive formula
\qq
F_{2M}\ =\ \sum
\limits_{{\rm partitions}\atop\Pi{\ \rm of}\ \CP_{2M}}
\bigotimes\limits_{\pi\in\Pi}F^c_\pi
\label{cOn}
\qqq
in the, hopefully, self-explanatory notation.
The connected \s$2M$-correlator \s\s$<T^{\m\otimes 2M}>^c\s\s$
is given by Eq.\s\s(\ref{pairi}) with \s$F_{2M}^c\s$ replacing
\s$F_{2M}\s$ on the right hand side.
Note that, by virtue of the definition (\ref{CM})
of \s$\CM_{N}\s$,
for each partition \s$\Pi\s$ of \s$\CP_{2M}\s$,
we may write
\qq
\CM_{2M}\ =\ \sum\limits_{\pi\in\Pi}\CM_\pi
\s-\s\sum\limits_{\{\pi,\m\pi'\}\subset\Pi
}\CL_{\pi,\pi'}
\label{eapa}
\qqq
where \s$\CM_\pi\s$ is \s$\CM_{2|\pi|}\s$ acting
on the tensor factors of \s$(\NR^d)^{\otimes\m{2M}}\s$
labeled by the pairs of \s$\pi\s$ and
\qq
\CL_{\pi,\pi'}\ =\ \sum\limits_{n\in\pi\atop n'\in\pi'}
(\m\CK)_{n,\m n'}\ .
\qqq
The basic equation for the connected single $2$-particle
channel correlation functions \s$F^c_{2M}\s$ is given by
\vs 1cm

\no{\bf Proposition 3}.\ \ For \s$M>1\s$,
\qq
\CM_{2M}\s\s F^c_{2M}\ =\ \sum\limits_{\Pi=\{\pi,\pi'\}
}\CL_{\pi,\pi'}
\s\m(\m F^c_\pi\otimes F^c_{\pi'}\m)\ .
\label{eqcon}
\qqq
\vs 0.7cm

\no Note the connected character of the right hand side.
Let us prove this relation. From the definition (\ref{1chan})
of the single channel function,
\qq
\CM_{2M}\s\s F_{2M}
\ =\ \sum\limits_{\Pi=\{\pi,\m\pi'\}\atop|\pi|=1}
\CC_{\pi}\otimes F_{\pi'}\ .
\label{SCH}
\qqq
On the other hand, using Eqs.\s\s(\ref{cOn})
and (\ref{eapa}), we obtain
\qq
\CM_{2M}\s\s F^c_{2M}\s&=&\s\CM_{2M}\s\s F_{2M}
\s-\s\sum\limits_{\Pi\atop|\Pi|>1}\CM_{2M}
\bigotimes\limits_{\pi\in\Pi}F^c_\pi\cr
&=&\s\CM_{2M}\s\s F_{2M}\s-\s\sum\limits_{\Pi\atop|\Pi|>1}
\sum\limits_{\pi\in\Pi}(\m\CM_\pi\s F^c_\pi\m)
\bigotimes\s(\bigotimes\limits_{\pi'\in\Pi\atop\pi'\not=\pi}
F^c_{\pi'}\s)\cr
&+&\s\sum\limits_{\Pi\atop|\Pi|>1}
\sum\limits_{\{\pi,\m\pi'\}\subset\Pi
}(\m\CL_{\pi,\pi'}(F^c_\pi\otimes F^c_{\pi'})\m)
\bigotimes\s(\hs{-0.1cm}
\bigotimes_{\pi''\not=\pi,\pi'}F^c_{\pi''}\s)\ .
\qqq
Notice that on the right hand side, in virtue
of Eqs. (\ref{SCH}) and (\ref{2pst}),
the $1^{\m\rm st}$
term cancels with
the contribution to the $2^{\m\rm nd}$ one from \s$\pi\s$
consisting of a single pair.
Rewriting \s$\CM_\pi\s F^c_\pi\s$
for the \s$|\pi|>1\s$ contributions to the
$2^{\m\rm nd}$ term
using Eq.\s\s(\ref{eqcon}) as the inductive hypothesis,
we observe that all these contribution are cancelled
by the $3^{\m\rm rd}$ term. What is left
from the $3^{\m\rm rd}$ term after the cancellations
gives exactly the right hand side of Eq.\s\s(\ref{eqcon}).
\vs 0.4cm

Although we worked directly with the evolution
equation (\ref{FDE}) in order to obtain expressions
(\ref{2pst}) and (\ref{1chan}), one may obtain
them as well using the MSR formalism which gives
\qq
< T^{\otimes\m2M}>\s=\s{_1\over^\CN}\s\int T^{\otimes\m2M}
\s\m\ee^{\m i\m(\m A\m,\s(\da_t+\nu\m\alpha+\beta)\m T\m)\m-
\m{_1\over^2}\m(A,\m\CC\m A)\m
-\m{_1\over^2}\m
(\beta,\m\CB^{-1}\m\beta)}\s\s DA\s\s DT\s\s D\beta\ .
\qqq
Performing the Gaussian \s$A$- and \s$T$-integrals, we obtain
\qq
< T^{\otimes\m2M}>\s=\s{_1\over^\CN}\hs{-0.4cm}
\sum\limits_{{\rm parings}\atop{{\rm of}\ \{1,\dots,2M\}}}
\hs{-0.35cm}\int\prod\limits_{\rm pairs}
(\da_t+\nu\m\alpha+\beta)^{-1}\m\CC
\m(-\da_t+\nu\m\alpha-\beta)^{-1}\s\s\ee^{\m-\m{_1\over^2}\m
(\beta,\m\CB^{-1}\m\beta)}\s\m D\beta\s.\hs{0.5cm}
\label{cos}
\qqq
If we expand
\qq
(\da_t+\nu\m\alpha+\beta)^{-1}\s=\s\sum\limits_{m=0}^\infty
(-1)^{m}\s(\da_t+\nu\m\alpha)^{-1}\left(\beta\s
(\da_t+\nu\m\alpha)^{-1}\right)^m
\qqq
and use the fact that the kernel of \s$(\da_t+\nu\m\alpha)^{-1}\s$
is equal to \s$\theta(t-t')\s\m\ee^{\m-\m(t-t')\m\nu\alpha}\s$,
\s the \s$\beta$-integral in (\ref{cos}) reduces to the
previous calculation and it reproduces the expressions
(\ref{2pst}) for the $2$-point function or (\ref{pairi})
and (\ref{1chan})
for the $2M$-correlator.
\vs 0.7cm

\subsection{Eddy diffusion versus super-diffusion}
\vskip 0.3cm

The exact solution for the PS correlation
functions obtained above in the case of finitely many
degrees of freedom still makes sense for the
infinite-dimensional case of the evolution equation
(\ref{PS}). Instead of starting with this equation,
and extending the previous analysis from the stochastic
ODE case to the stochastic PDE one,
which would require additional work,
we shall directly study the infinite-dimensional
version of the exact solution (\ref{pairi}), (\ref{1chan})
involving operators rather (generally unbounded) than
finite matrices. The analysis of the expressions
for the PS correlation functions
obtained this way is more complicated than in
finite dimensions, but also more interesting.
\vs 0.3cm

The covariance \s$\CB\s$ of the Gaussian process
\s$\beta=\Nv\cdot\nabla\s$ and the operator \s$B\s$
related to the expectation of \s$\beta^2\s$, \s see
Eqs.\s\s(\ref{B}) and (\ref{B1}), are given by
the kernels
\qq
&&\CB(\Nx_1,\Ny_1;\Nx_2,\Ny_2)\ =\ \CD^{ij}(\Nx_1-\Nx_2)\s
\s\m\da_{i}\m
\delta(\Nx_1-\Ny_1)\s\s\m\da_j\m\delta(\Nx_2-\Ny_2)\ ,\cr
&&B(\Nx,\Ny)\ =\ -\hf\m\int\CB(\Nx,\Nz;\Nz,\Ny)\s d^3\Nz
\s=\s-\hf\CD^{ij}(0)\m\s\da_i
\da_j\s\delta(\Nx-\Ny)\ .
\qqq
Since \s\s$\CD^{ij}(0)\m=\m{_1\over^3}\s
\delta^{ij}\s\m\CD^{ll}(0)\s$,
\s\s$B\s=\s-{1\over6}\s\CD^{ll}(0)\s\m\Delta\s\s$
and, consequently,
\qq
\CM_1\s\equiv\s\nu\s\alpha\s+\s B\s=\s-
(\nu+{_1\over^6}\m\CD^{ll}(0))\m\s\Delta\s\equiv
\s-\nu_{\rm eff}\s\m\Delta\ .
\qqq
It follows from Eq.\s\s(\ref{decay}) with \s$N=1\s$
that, in the absence of the sources,
the expectation value of the scalar diffuses
\qq
<\s T(t,\Nx)>&=&\int\ee^{\m-\m(t-t_0)\m\CM_1}(\Nx,\Ny)\s\s
T(t_0,\Ny)\s\s d^3\Ny\cr
&=&\int\ee^{\m(t-t_0)\m\nu_{\rm eff}\m\Delta}(\Nx,\Ny)\s\s
T(t_0,\Ny)\s\s d^3\Ny
\qqq
with the effective diffusion constant \s$\nu_{\rm eff}\s$
composed of the molecular diffusivity \s$\nu\s$
and the {\bf eddy diffusivity}
\s${1\over6}\m\CD^{ll}(0)\s$. \s
For small \s$m\s$, \s the eddy diffusivity dominates
and the diffusion is driven by the large distance scales
(recall from Eq.\s\s(\ref{div})
that \s$\CD^{ll}(0)=\CO(D_0\s m^{-\kappa})\s$)\m.
\vs 0.3cm

Operator \s$\CK\s$ with kernel
\s\s$\CK(\Nx_1,\Nx_2;\Ny_1,\Ny_2)
=\CB(\Nx_1,\Ny_1;\Nx_2,\Ny_2)\s\s$ acts on functions
\s$g\s$ of six variables by
\qq
(\m\CK\s g)(\Nx_1,\Nx_2)\s=\s
\CD^{ij}(\Nx_1-\Nx_2)
\s\s\da_{x_1^i}\m\da_{x_2^j}\s g(\Nx_1,\Nx_2)\ .
\qqq
so that for the \s$N$-body operators \s$\CM_N\s$ given by formula
(\ref{CM}) we obtain
\qq
\CM_N\s&=\s&-\sum\limits_{n=1}^N
(\nu\s\Delta_{\Nx_n}+\hf\m\CD^{ij}(0)\s\da_{x_n^i}
\da_{x_n^j})\s-\s\sum\limits_{n<n'}\CD^{ij}(\Nx_n-\Nx_{n'})
\s\da_{x_n^i}\da_{x_{n'}^j}\ \cr
&=\s&-\sum\limits_{n=1}^N\nu_{\rm eff}\s\Delta_{\Nx_n}
\s-\s\sum\limits_{n<n'}\CD^{ij}(\Nx_n-\Nx_{n'})
\s\da_{x_n^i}\da_{x_{n'}^j}\ .
\label{cmn}
\qqq
In particular,
\qq
\CM_2\s=\s-\nu\m(\Delta_1+\Delta_2)\m
-\m\hf\s\CD^{ij}(0)\s\da_{x_1^i}\da_{x_1^j}
-\m\hf\s\CD^{ij}(0)\s\da_{x_2^i}\da_{x_2^j}
-\m\CD^{ij}(\Nx_1-\Nx_{2})
\s\da_{x_1^i}\da_{x_{2}^j}\ .\ \ \ \
\qqq
Note that \s$\CM_2\s$ commutes with (three-dimensional)
translations and in the action on translation-invariant
functions of \s$\Nx_1-\Nx_2\equiv\Nx\s$
reduces to
\qq
\CM_2\s=\s-2\m\nu\s\Delta\s-\s\tilde\CD^{ij}(\Nx)
\s\da_{i}\da_{j}\ .
\qqq
Since \s$\tilde\CD^{ij}(\Nx)\equiv\CD^{ij}(0)-\CD^{ij}(\Nx)\s$
has an \s$m\to 0\s$ limit, so does the operator
\s$\CM_2\s$ in the action on translation-invariant
functions and when \s$\nu\to 0\s$, it becomes
a singular elliptic operator
\qq
\CM_2^{\rm sc}\ =\ -\m D_1\left((2+\kappa)
\s\delta^{ij}\s|x|^{\kappa}\s-
\s\kappa
\s x^i\m x^j\s|x|^{\kappa-2}\right)\da_{i}\da_{j}\ ,
\label{M2sc}
\qqq
where \s\s$D_1\s\equiv\s{{\Gamma((2-\kappa)/2)}
\over{2^{2+\kappa}\pi^{3/2}
\kappa(3+\kappa)}}\s\m D_0\s$, \s see Eq.\s\s(\ref{td}).
The dimension of \s$D_1\s$, as that of \s$D_0\s$, is equal to
\s$length^{2-\kappa}\over time\s$.
\vs 0.3cm

Similarly for \s$N>2\s$,
in the action
on the translationally invariant sector,
\qq
\CM_N\s=\s-\nu\sum\limits_{n=1}^N\Delta_{\Nx_n}\s+\s
\sum\limits_{n<n'}\tilde\CD^{ij}(\Nx_n-\Nx_{n'})\s\da_{x_n^i}
\da_{x_{n'}^j}
\label{448}
\qqq
and has an \s$m\to0\s$ and \s$\nu\to0\s$ limit
which is a singular elliptic operator \s$\CM_N^{\rm sc}\s$.
This has important consequences. Although
the time evolution given by Eq.\s\s(\ref{decay})
of fixed initial conditions is dominated for small
\s$m\s$ by the eddy diffusivity diverging as \s$m\to0\s$,
the time evolution governed by Eq.\s\s(\ref{decay1})
for random initial data with
a translationally invariant distribution
behaves well in the limit
when \s$m\to0\s$ and \s$\nu\to 0\s$. \s In particular,
it follows from the scaling properties of operators
\s$\CM_N^{\rm sc}\s$ that, if the translationally
invariant initial moments \s$<\prod_n T(t_0,\Nx_n)>\s$
are fast decaying in the
difference variables (weakly correlated initial data)
then, for \s$m=0\s$, \s the rescaled correlations
\qq
\lambda^{3N\over 2(2-\kappa)}\s<\prod\limits_{n=1}^N
T(\lambda t\m,\s\lambda^{1\over2-\kappa}\Nx_n)>
\qqq
have a non-trivial limit when \s$\lambda\to\infty\s$.
In other words, the equal-time correlators
decay {\bf super-diffusively} in the absence of sources,
like \s$t^{-{3N\over 2(2-\kappa)}}\s$,
\s i.\s e. faster than diffusively
(like \s$t^{-{3N\over 4}}\s$)\m.
\s Similar behavior in a simplified
model of passive advection was discussed in \cite{Majda}.
\vs 0.4cm

Let us look closer at the operator \s$\CM_2\s$ in the
translationally invariant sector. Since the limit
\s$m\to0\s$ presents no difficulty in that case,
we may directly work in the \s$m=0\s$ case, i.\s e. with
operator \s$\CM_2=-2\m\nu\s\Delta\s+\s\CM_2^{\rm sc}\s$, \s see
Eq.\s\s(\ref{M2sc}). \s In the radial variables
\qq
\CM_2\ =\ -{_{2}\over^{r^2}}\s\da_r
\s(\m\nu\s r^2\s+\s D_1\s r^{2+\kappa}\m)\m\s\da_r\s+
\s{_{l(l+1)}\over^{r^2}}\s(\m 2\m\nu\s+\s D_1\m(2+\kappa)\s
r^{\kappa-2}\m)\ ,
\qqq
where \s$l\s$ is the angular momentum \s(\s$\CM_2\s$ commutes
with rotations).
As for the resolvents \s$(i\omega+\CM_2)^{-1}\s$, they may be
explicitly computed in the translationally and rotationally
invariant ($\m l=0\m$) sector when \s$\nu\to0\s$.
In particular, one obtains for the kernel
\qq
(i\omega+\CM_2^{\rm sc})^{-1}(\Nx,0)
\ =\ {\omega}^{1+\kappa\over 2-\kappa}
\s\m f(\omega^{1\over2}|\Nx|^{{2-\kappa\over 2}})
\label{hank}
\qqq
with an exponentially decaying function \s$f\s$,
\s$f(y)\sim y^{-{2(1+\kappa)\over 2-\kappa}}\s$
for small \s$y\s$. \s$f\s$ may be expressed in terms of
the Hankel functions. This result has an important
implication. Recall that in finite dimensions the invariant
measures for the stochastic evolution of the initial data
were characterized by the symmetry generated by
rotations \s$\ee^{\m\beta^I}\s$. \s Suppose that,
as for a finite
number of degrees of freedom, the evolution
(\ref{decay1}) preserves in the limit
\s$\nu\to0\s$ the white noise
Gibbs measure \s$\sim\ \ee^{\m-{\rm const}.\m\|T\|^2}\s\s
DT\s$ with the $2$-point function
\qq
< T(\Nx)\s\m T(\Ny)>\ \sim\s\delta(\Nx-\Ny)\ .
\qqq
This would imply that
\qq
\int g(\Nx)\s\s(i\omega+\CM_2^{\rm sc})^{-1}(\Nx,0)\s\s
d^3\Nx\s=\s{1\over{i\omega}}\s\m g(0)
\label{prev}
\qqq
for any test function \s$g\s$, which is in contradiction
with Eq.\s\s(\ref{hank}). Taking Fourier transforms
in space and time of the latter, one
obtains
\qq
\int\ee^{-i\Nk\cdot\Nx}\ \ee^{-t\s\CM_2^{\rm sc}}(\Nx,0)
\s\s d^3\Nx
\s\s=\s\s \rho(\m t^{1\over 2-\kappa}\m\vert\Nk\vert\m)\ ,
\label{poro}
\qqq
with \s$\rho(0)=1\s$ and \s$\rho\s$ decaying at infinity,
instead of a constant as would be implied by Eq.\s\s(\ref{prev}).
One should compare the relation (\ref{poro}) to the relation
\qq
\int\ee^{-i\Nk\cdot\Nx}\ \ee^{-t\m\Delta}(\Nx,0)
\s\s d^3\Nx
\s=\s\ee^{-t\m\vert{\bf k}\vert^2}
\qqq
describing the diffusive decay of the energy spectrum.
For example, Eq.\s\s(\ref{poro}) implies that
for homogeneous, fast decaying initial data
$2$-point function,
\qq
\int<T(t,0)\s\s T(t,\Nx)>\s\s\Nx^2\s\s d^3\Nx\ \
\sim\ \ t^{2\over 2-\kappa}
\qqq
instead of being proportional to \s$t\s$:
again a super-diffusive behavior. The reason of the failure
of the Gibbs measure to be invariant under the evolution
(\ref{decay1}) for infinitely many degrees of freedom
is the energy cascade towards degrees of freedom with
smaller and smaller wavenumbers, see below.
This failure shows that, for the infinite number of degrees
of freedom, the symmetry group generated
by the exponentials \s$\ee^{\m\beta^I}\s$, \s which
for \s$\beta=\Nv\cdot\nabla\s$ is the group of volume
preserving diffeomorphisms of \s$\NR^3\s$, \s is broken.
It is not excluded, however, that this symmetry may still
be useful in analysis of the model.
\vs 0.7cm

\subsection{Quasi-Lagrangian approach to passive scalar}
\vskip 0.3cm

PS constitutes a perfect model to test the
quasi-Lagrangian ideas mentioned at the end of Section 4.
The scalar density \s$T'(t,\Nx)\m\equiv\m T(t,\Nx+\Nx(t))\s$
viewed from a point moving in the
fluid satisfies the equation
\qq
\da_tT'\s+\s(\Nv'-\Nv'_0)T'\s-\s\nu\Delta T'\s=\s f'
\qqq
where \s$\Nv'(t,\Nx)\m\equiv\m \Nv(t,\Nx+\Nx(t))\s$
and \s$f'(t,\Nx)\m\equiv\m f(t,\Nx+\Nx(t))\s$
in the notation of Section 4. It is again a stochastic
equation of the type (\ref{FDE}) with
\s$\beta=(\Nv'-\Nv'_0)\cdot\nabla\s$.
\s One may show that for Gaussian
homogeneous \s$\Nv(t,\Nx)\s$ and \s$f(t,\Nx)\s$, \s white
noise in time,  the distributions of the random
fields \s$\Nv'(t,\Nx)\s$ and
\s$f'(t,\Nx)\s$ obtained by the quasi-Lagrangian
transformation become the same as those of \s$\Nv(t,\Nx)\s$
and \s$f(t,\Nx)\s$ asymptotically
for large times. For
\s$\Nv'(t,\Nx)\s$ and \s$\Nf'(t,\Nx)\s$ distributed
as \s$\Nv(t,\Nx)\s$ and \s$f(t,\Nx)\s$ before, we
may produce the exact solution for
the correlation functions of \s$T'\s$ by replacing
the operators \s$\CM_N\s$ by new operators \s$\CM'_N\s$
obtained by changing the velocity correlation function
(\ref{42}) to
\qq
<({v'}^{\m i}(t,\Nx)-{v'}^{\m i}(t,0))
\m\s({v'}^{\m j}(t',\Nx')-{v'}^{\m j}(t',0))>&&\cr\cr
&&\hs{-6cm}=\s\s\delta(t-t')\s\left(D^{ij}
(\Nx-\Nx')-D^{ij}(\Nx)-D^{ij}(\Nx')
+D^{ij}(0)\right)\cr
&&\hs{-6cm}=\s\s-\s\delta(t-t')\s
\left(\tilde D^{ij}(\Nx-\Nx')-\tilde D^{ij}(\Nx)
-\tilde D^{ij}(\Nx')\right)\ .
\label{corR}
\qqq
Notice that only the subtracted functions \s$\tilde D^{ij}\s$
with the regular \s$m\to 0\s$ limit
enter the correlator (\ref{corR}).
One obtains
\qq
\CM'_N\s&=&\s\nu\sum\limits_{n=1}^N\Delta_{\Nx_n}
\s-\sum\limits_{n=1}^N\tilde D^{ij}(\Nx_n)\s\m\da_{x^i_n}
\s\da_{x_{n}^j}\ \s\s\cr
&+&\s\sum\limits_{n<n'}
\left(\tilde D^{ij}(\Nx_n-\Nx_{n'})-\tilde D^{ij}(\Nx_n)
-\tilde D^{ij}(\Nx_{n'})\right)
\s\m\da_{x^i_n}\s\da_{x^j_{n'}}\ .
\label{cmn'}
\qqq
The crucial observation is that in the action on
translationally invariant functions operators \s$\CM_N'\s$
and \s$\CM_N\s$, the latter given by Eq.\s\s(\ref{448}),
coincide (the terms with \s$\tilde D^{ij}(\Nx_n)\s$
cancel). Hence for the PS, the stationary
equal time quasi-Lagrangian correlators will be equal
to the original (Eulerian) ones for the homogeneous
sources. This is not the case for the non-equal time
correlators which involve the heat operators
\s$\ee^{-t\CM_N}\s$
\s(\s$\ee^{-t\CM'_N}\s$) \s in
the non-translationally invariant sector.
The latter are governed by the eddy diffusivity, i.\s e. by
the integral scale in the Eulerian case
but show a universal behavior in the inertial range
in the quasi-Lagrangian description.
The quasi-Lagrangian picture is also better behaved from
the point of view of the standard perturbative approach
which consists of treating the $2$-particle terms
in \s$\CM_N\s$ as a perturbation
of the $1$-particle ones, see Eq.\s\s(\ref{CM}).
In the Eulerian approach, the unperturbed terms are
dominated by the diverging eddy diffusivity
whereas in the quasi-Lagrangian
one the $1$-particle terms have a regular limit when
\s$m\to 0\s$, \s compare Eqs.\s\s(\ref{cmn}) and (\ref{cmn'}).
Thus the quasi-Lagrangian perturbation expansion
seems to be a better tool for the analysis of the
behavior of the model.
\vs 1cm

\nsection{Steady state of the passive scalar}
\subsection{Universality of the $2$-point function}
\vskip 0.5cm

Let us pass to the study of the stationary
state $2$-point function \s$< T(\Nx_1)\s\m T(\Nx_2)>\ \equiv
\s F_2(|\Nx_1-\Nx_2|)\s$
in the presence of the random sources with
a rotationally invariant covariance \s$\CC_L=\CC(\m\cdot\m/L)\s$,
\s see Eq.\s\s(\ref{force}).
\s$F_2\s$ satisfies the differential equation
\qq
(\m\CM_2\s F_2\m)(r)\ =\ -{_{2}\over^{r^2}}\s\da_r
\s(\m\nu\s r^2\s+\s D_1\s r^{2+\kappa}\m)\m\s\da_r\s\m F_2(r)
\ =\ \CC_L(r)
\label{wisi}
\qqq
which is implied by Eq.\s\s(\ref{2pst}). Integrating once,
we obtain
\qq
\da_r\s F_2(r)\ =\ -\hf\s{\smallint\limits_0^r
\CC(\rho/L)\s\rho^2\s d\rho\over
\nu\s r^2\s+\s D_1\s r^{2+\kappa}}\ ,
\label{F20pr}
\qqq
where the constant of integration has been chosen
so as to obtain \s$\da_r\m F_2\s$ vanishing
at zero as required if \s$F_2\s$ is to be
a function on \s$\NR^3\s$
regular at zero. One more integral
gives
\qq
F_2(r)\ =\
\hf\s\int\limits_r^\infty\m{\smallint\limits_0^\rho
\CC(\rho'/L)\s{\rho'}^2\s d\rho'\over
\nu\s \rho^2\s+\s D_1\s\rho^{\m2+\kappa}
}\s\s\m d\rho\ ,
\label{m0}
\qqq
where the choice of integration constant
assures that \s$F_2\s$ decays at infinity.
We want to study the behavior of \s$F_2(r)\s$
when the integral scale \s$L\to\infty\s$. \s To
this end, let us rewrite the $\rho$-integral
from \s$r\s$ to \s$\infty\s$ as a difference
of the integral from \s$0\s$ to \s$\infty\s$
and from \s$0\s$ to \s$r\s$:
\qq
F_2(r)\ =\
\hf\s L^{2-\kappa}\int\limits_0^\infty\m{\smallint\limits_0^\rho
\CC(\rho')\s{\rho'}^2\s d\rho'\over
L^{-2-\kappa}\s\nu\s \rho^2\s+\s D_1\s\rho^{\m2+\kappa}
}\s\s\m d\rho\ -\
\hf\s\int\limits_0^r\m{\smallint\limits_0^\rho
\CC(\rho'/L)\s{\rho'}^2\s d\rho'\over
\nu\s \rho^2\s+\s D_1\s\rho^{\m2+\kappa}
}\s\s\m d\rho\ ,
\label{m1}
\qqq
The first of the integrals is
a \s$\CC$-dependent constant diverging
when \s$L\to\infty\s$. On the other hand,
the integral from \s$0\s$
to \s$r\s$ has a limit when \s$L\to\infty\s$
equal to
\qq
-{_1\over^6}\m\s\CC(0)\int\limits_0^r{{\rho^3\s d\rho}\over
{\nu\m\rho^2+D_1\m\rho^{2+\kappa}}}
\qqq
which is universal (i.e. depends only on \s$\CC(0)\s$
which, as we shall see in the next Section, equals
twice the energy dissipation rate \s$\epsilon\s$ of
the scalar). Note that the non-universal term (a constant)
is annihilated by \s$\CM_2\s$. This has to be so
if Eq.\s\s(\ref{wisi}) is to be satisfied. The
right hand side becomes universal in the limit \s$L\to\infty\s$
so all non-universal terms in \s$F_2(r)\s$ surviving
in this limit have to be annihilated by \s$\CM_2\s$.
We shall see this general mechanism limiting possible
non-universal terms also for the higher point functions.
The constant term of \s$F_2\s$ is automatically subtracted
in the $2^{\m\rm nd}$ structure function:
\qq
S_2(r)\s=\s2\m(\m F_2(0)-F_2(r)\m)\ =\
{_2\over^3}\s\epsilon\int\limits_0^r{{\rho^3\s d\rho}\over
{\nu\m\rho^2+D_1\m\rho^{2+\kappa}}}
\qqq
where the last equality holds for infinite \s$L\s$.
\s In the \s$\nu\to0\s$ limit,
\qq
S_2(r)\ =\ {_{2\s\epsilon}\over^{3\m(2-\kappa)\m D_1}}
\  r^{2-\kappa}\
\label{IR2}
\qqq
and the same universal result holds approximately in
the whole inertial range \s$\eta\ll r\ll L\s$
where at the Kolmogorov scale
\s$\eta=\CO(\m(\nu/D_1)^{1/\kappa})\s$ the term
\s$\nu\s r^2\s$ becomes comparable to \s$D_1\s\rho^{2+\kappa}\s$.
The $2^{\m\rm nd}$ structure
function exponent \s$\zeta_2=2-\kappa\s$.
\vs 0.4cm

It is instructive to repeat the same analysis for
anisotropic sources, i.\s e. without
assuming that the correlation \s$\CC_L\s$ is
rotationally invariant. We shall have then to solve
the equation
\qq
\CM_2\s\m F_2\s =\s \CC_L
\label{equ}
\qqq
in other angular momentum sectors. This is an
easy exercise \cite{my} and one still obtains,
for \s$m\s$ and \s$\nu\s$ equal to zero,
the result
\qq
F_2(\Nx)\ -\
\ \gamma\s\s L^{\m 2-\kappa}
\ \
\smash{\mathop{\longrightarrow}
\limits_{L\to\infty}}\ \
-\ {_{\epsilon}\over^{3\m (2-\kappa)\m D_1}}
\s\s \vert\Nx\vert^{\m 2-\kappa}\ \equiv\ F_2^{\rm sc}(\Nx)\ ,
\label{limi}
\qqq
where \s$\gamma\s$ is a \s$\CC$-dependent constant.
In the case of a complex PS, however,
further subtractions are needed when \s$0<\kappa<1\s$.
\s For \s$F_2(\Nx)\s\equiv\s
\s<T(0)\s\m \overline{T(\Nx)}>\s$, \s one shows
then \cite{my} that
\qq
F_2(\Nx)\ -\
\ \gamma\s\s L^{\m 2-\kappa}
\ -\
\gamma_{i}\s x^i\s\s L^{\m 1-\kappa}
\ \
\smash{\mathop{\longrightarrow}
\limits_{L\to\infty}}\ \
-\ \s{_{\epsilon}\over^{3\m (2-\kappa)\m D_1}}
\s\s\vert\Nx\vert^{\m 2-\kappa}\ ,
\label{limi1}
\qqq
where real \s$\gamma\s$ and imaginary \s$\gamma_i\s$
are \s$\CC$-dependent constants. The \s$\gamma_i$-terms
are present only if \s$\CC\s$ has angular momentum $1$
components and this is impossible in the real
case where only even \s$l\s$'s appear\footnote{K.G. would
like to thank R. Kraichnan for pointing this out to
him}. Again, the non-universal terms are homogeneous
zero modes of \s$\CM_2\s$ which annihilates
constant and linear functions, in accordance with the general
argument. Similarly as the constant,
the linear terms appearing with the coefficient
\s$L^{\m1-\kappa}\s$ in the presence of anisotropic
sources for the complex scalar
drop out from the $2^{\m\rm nd}$
structure function \s$<|T(\Nx)-T(0)|^2
\hs{-0.05cm}>\s=\s 2\s F_2(0)-F_2(\Nx)-F_2(-\Nx)\s$
and do not spoil the inertial range universality
of the latter.
\vs 0.7cm

\subsection{Energy cascade}
\vskip 0.3cm

Let us look at the energy balance
of the scalar. The dynamical equation
(\ref{2poin}) takes the form
\qq
\da_t F_2(r)\ =\ -(\CM_2\s F_2)(r)\s+\s \CC_L(r)\ .
\label{dyeq}
\qqq
and reduces in the stationary state to
Eq.\s\s(\ref{wisi}). In particular, the mean energy
density \s$e\s\equiv\s\s<\hf T^2>\s$
satisfies
\qq
\da_t\m e\ =\s\hf\s
\da_t F_2(0)&=&-\hf\m(\CM_2\s F_2)(0)\s+
\s\hf\s \CC(0)\cr
&=& \m\nu\m(\Delta F_2)(0)
\s+\s\hf\s \CC(0)
\qqq
(only the \s$-2\m\nu\m\Delta\s$ term of \s$\CM_2\s$
contributes at \s$r=0\s$)\m.
\qq
\epsilon\ \equiv
\s-\m\nu\m(\Delta\m F_2)(0)\s=
\s-\m\nu\m{_1\over^{r^2}}\m\da_r\m r^2
\m\da_r\m F_2(r)|_{_{r=0}}
\label{dssp}
\qqq
is the mean dissipation rate and \s$\hf \m\CC(0)\s$
is the mean injection rate of
energy (both intensive).
In the steady state the energy
balance reduces to the equality
of the rates :
\qq
\epsilon\ =\s\hf\s\CC(0)\ .
\qqq
It is, of course, satisfied by
our solution for the steady state
$2$-point function.
\vs 0.3cm

We would like to see how the energy
is distributed
among different wavenumbers following
the schematic discussion of Section 1.2.
For the mean energy density of modes
with \s$|\Nk|\leq K\s$, \s$e_{\leq K}
\s\equiv\s\s<\hf\s T_{\leq K}^{\s\s2}>\m$,
\s we obtain
\qq
e_{\leq K}\ =\s\hf\s\smallint\limits_{|\Nk|\leq K}
\left(\smallint\ee^{\m-i\m\Nk
\cdot\Ny}\s\s F_2(\Ny)\s\s d^3\Ny\right)
\s{_{d^3\Nk}
\over^{(2\pi)^3}}
\ \equiv\
\smallint\limits_0^Ke(k)\s\m dk
\qqq
with the energy spectrum
\qq
e(k)\s\sim\s k^{-3+\kappa}
\qqq
in the steady state inertial range.
The dynamical equation (\ref{dyeq})
implies that
\qq
\da_t\s e_{\leq K}\ =\s-\epsilon_{\leq K}
\s+\s\varphi_{\leq K}\s-\s\pi_{K}\s\ ,
\label{enfl}
\qqq
where
\qq
\epsilon_{\leq K}&\equiv&
-\s\nu\smallint\limits_{|\Nk|
\leq K}\left(\smallint\ee^{- i\m\Nk
\cdot\Ny}\s\s\Delta\m F_2(\Ny)\s\m d^3\Ny\right)
\s{_{d^3\Nk}\over^{(2\pi)^3}}\cr\cr
&=&-\m{_2\over^\pi}\s\nu\m\smallint\limits_0^\infty
{_{\sin(Kr)\s-\s K\m r\s\cos(Kr)}\over^{r^3}}\s\s\s
\da_r\s r^2\s\da_r\s F_2(r)\s\s dr
\label{dissi}
\qqq
is the mean energy
dissipation rate in the modes with \s$|\Nk|\leq K\s$ and
\qq
\varphi_{\leq K}\ \equiv\s
\hf\s\smallint\limits_{|\Nk|\leq K}
\left(\smallint\ee^{-i\m\Nk\cdot\Ny}
\s\s\CC_L(\Ny)\s\s
d^3\Ny\right)\s\s{_{d^3\Nk}\over^{(2\pi)^3}}
\s=\s{_1\over^\pi}\m\smallint\limits_0^\infty
{_{\sin(Kr)\s-\s K\m r\s\cos(Kr)}\over^{r}}\s\s\s
\CC(r/L)\s\s dr\hs{0.5cm}
\label{inje}
\qqq
is the mean energy injection rate into
\s$|\Nk|\leq K\s$.
\qq
\pi_{K}\ \equiv\s
-\m{_{2\m D_1}\over^\pi}\m\smallint\limits_0^\infty
{_{\sin(Kr)\s-\s K\m r\s\cos(Kr)}\over^{r^3}}\s\
\da_r\s\m r^{2+\kappa}\m\s\da_r\s F_2(r)\s\s dr
\qqq
comes from the contributions of
\s$(\CM_2^{\rm sc}\s F_2)(r)\s$
to (\ref{dyeq}). In view of the
relation (\ref{enfl}), it is
natural to interpret \s$\pi_{K}\s$
as the mean energy flux out of the
modes with \s$|\Nk|\leq K\s$, compare Eq.\s\s(\ref{DEB})
in Sect.\s\s2. In particular,
in the forced steady state, we obtain
the energy balance equation
\qq
\varphi_{\leq K}\ =\s\epsilon_{\leq K}\s+
\s\pi_{K}\ ,
\label{enst}
\qqq
compare Eq.\s\s(\ref{MDEB}).
\vs 0.3cm

With the Fourier transform of \s$\CC(\m\cdot\m/L)\s$
concentrated around the zero wavenumbers,
the energy injection is essentially limited
to small wavenumbers and
the mean injection rate
\s$\varphi_{\leq K}\s$ is approximately
constant for \s$K\gg L^{-1}\s$ and equal
to its \s$K=\infty\s$ value \s$\hf\s\CC(0)\s$.
\s On the other hand, the
mean dissipation rate \s$\epsilon_{\leq K}\s$
may be easily showed for \s$\kappa>1\s$ to satisfy
\qq
\epsilon_{\leq K}\ \leq\
\CO(\m{{\nu\m\epsilon}\over D_1}\s\m K^{\m\kappa}\m)\ .
\label{dissbd}
\qqq
Indeed, it follows from Eq.\s\s(\ref{F20pr}) that
\s\m$\vert\da_r r^2\da_r
F_2(r)\vert\leq{\rm const}.\s\s D_1^{\m-1}\s r^{2-\kappa}\s$.
Estimating
\qq
\vert\sin(Kr)-Kr\cos(Kr)\vert\s\leq\s\cases{
{\rm const}.\s\s(Kr)^3\hs{0.6cm}{\rm for}\hs{0.6cm}
r\leq K^{-1}\ ,\cr
{\rm const}.\s\s Kr\ \s\hs{0.6cm}{\rm for}\hs{0.6cm}
r\geq K^{-1}\ ,}
\qqq
we obtain
\qq
\epsilon_{\leq K}\ \leq{\rm const}.\s\s{{\nu\m\epsilon}\over D_1}
\left(K^3\smallint\limits_0^{1/K}r^{2-\kappa}\s dr
\s+\s K\smallint\limits_{1/K}^\infty r^{-\kappa}\s dr\right)
\qqq
which implies the desired
bound (\ref{dissbd}) for \s$\kappa>1\s$.
\s For \s$\kappa\leq 1\s$, \s a
better estimate for \s$\vert\da_r
r^2\da_r F_2(r)\vert\s$ for large
\s$r\s$ using the decay of \s$\CC\s$
implies a somewhat worse bound
\qq
\epsilon_{\leq K}\ \leq\
\CO(\m{\nu\m\epsilon\over D_1}\m\s K^{\kappa}\m)\s
+\s\CO(\m{\nu\m\epsilon\s
L^{1-\kappa}\over D_1}\s\m K\m)\ .
\label{disbd}
\qqq
As a consequence, for \s$\kappa>1\s$, the mean dissipation rate
\s$\epsilon_{\leq K}\s$ stays very
small relative to \s$\epsilon\s$ as long as \s$K\ll  \eta^{-1}\s$
where \s$\eta=\CO(\m(\nu/D_1)^{1/\kappa})\s$
is the Kolmogorov scale. For \s$\kappa\leq 1\s$ this holds
at least for \s$K\ll \nu^{-1}\s D_1\s L^{\kappa-1}\s$.
\s The energy balance equation
(\ref{enst}), implies then that, for \s$\kappa>1\s$,
\s$\s L^{-1}\ll K\ll \eta^{-1}\s$,
\s the mean energy flux \s$\pi_{K}\s$
is almost equal to the mean injection rate
\s$\varphi_{\leq K}\s$
and hence almost constant and equal to
\s$\epsilon\s$ in the inertial range.
For \s$\kappa\geq 1\s$ the same is true at least for
\s$\s L^{-1}\ll K\ll \nu^{-1}\s D_1\s L^{\kappa-1}\s$. \s
This confirms an intuitive picture
of the non-dissipative energy cascade
through the decreasing
distance scales from the integral scale
\s$L\s$ where the energy is injected
to the Kolmogorov scale \s$\eta\s$ at least
for \s$\kappa>1\s$. At the end the energy
is dissipated on distances
shorter than \s$\eta\s$. \s What happens
for \s$\kappa\leq 1\s$ and \s$K\s$  between
\s$\CO(\nu^{-1}\s D_1\s L^{\kappa-1})\s$
and \s$\CO(\eta^{-1})\s$ remains to be understood.
\vs 0.7cm

\subsection{Gaussian case}
\vs 0.3cm

It will be instructive to look at the case when
\s$\kappa=0\s$ with
\qq
\tilde D^{ij}\s=\s 2\s D_1\s\m\delta^{ij}\ .
\qqq
(Note that having finite \s$D_1\s$ requires
infinitesimal \s$D_0\s$ when \s$\kappa\to 0\s$ in order
to renormalize the ultraviolet divergence
in the momentum-space integral
(\ref{43})\m; \s$D_0\s$ will never show up below.)
Although quite trivial, the case \s$\kappa=0\s$ appears
to be instructive and will serve below as the
departure point for an expansion in powers of \s$\kappa\s$.
\s We immediately obtain for the \s$\kappa=0\s$ operators\s:
\qq
\CM_{2}^{\rm sc}&=&2\s D_1\s\nabla_{\Nx_1}\cdot\nabla
_{\Nx_2}\s=\s-\m 2\s D_1\s\Delta_\NX\ ,\label{47}\\\cr
\CM_{4}^{\rm sc}&=&2\s D_1\sum\limits_{1\leq
n<n'\leq 4}\nabla_{\Nx_n}
\cdot\nabla_{\Nx_{n'}}\cr\cr
&=&-2\s D_1\m\left(\Delta_\NX\s+\s\Delta_\NY
\s+\s\Delta_\NZ\s-\s\nabla_\NX\cdot\nabla_\NY
\s-\s\nabla_\NY\cdot\nabla_\NZ\right)
\label{471}
\qqq
in the difference variables \s\s$\NX\equiv\Nx_1-\Nx_2\m,\
\NY\equiv\Nx_2-\Nx_3\m,\m\ \NZ=\Nx_3-\Nx_4\s$.
The \s$2$-point function
\qq
F_2(\Nx)\s=\s({\CM_2^{\rm sc}}^{^{-1}}\m \CC_L\m)(\Nx)
\s=\s{_1\over^{2\m D_1}}\int\ee^{\m i\m\Nk\cdot\Nx}
\s\s|\Nk|^{-2}\s\s\hat\CC_L(\Nk)\s\s {_{d^3\Nk}
\over^{(2\pi)^3}}\ ,
\label{2fc}
\qqq
where
\qq
\hat\CC_L(\Nk)\s=\s\int\ee^{-i\m\Nk\cdot\Nx}\s\s\CC(\Nx/L)\s\s d^3\Nx
\s=\s L^3\s\m\hat\CC(L\Nk)\ .
\qqq
Hence
\qq
F_2(\Nx)\s=\s{_1\over^{2\m D_1}}\s\s L^2\int
\ee^{\m i\m\Nk\cdot\Nx/L}\s\s|\Nk|^{-2}\s\m\hat\CC(\Nk)
\s\s{_{d^3\Nk}\over^{(2\pi)^3}}\ .
\label{seco}
\qqq
The large \s$L\s$ behavior may be obtained by
expanding \s$\ee^{\m i\m\Nk\cdot\Nx/L}\s$ to the second order\s:
\qq
F_2(\Nx)&=&{_1\over^{2\m D_1}}\s\s L^2\int|\Nk|^{-2}
\s\s\hat\CC(\Nk)\s\s{_{d^3\Nk}\over^{(2\pi)^3}}\s+\s
{_1\over^{2\m D_1}}\s\s L\s x^j\int
i\m k_j\m|\Nk|^{-2}\s\m\hat\CC(\Nk)\s\s
{_{d^3\Nk}\over^{(2\pi)^3}}\cr
&-&{_1\over^{4\m D_1}}\s\s x^i\m x^j\int
k_i\m k_j\m|\Nk|^{-2}\s\s\hat\CC(k)\s\s
{_{d^3\Nk}\over^{(2\pi)^3}}\s+\s\CO(L^{-1})\cr\cr
&=&{_1\over^{2\m D_1}}\s\s(\m(-\Delta)^{-1}\s\CC\m)(0)
\s+\s{_1\over^{2\m D_1}}\s\s(\m\da_i\m(-\Delta)^{-1}\s\CC\m)(0)
\s\s x^i\cr
&+&{_1\over^{4\m D_1}}\s\s(\m(\m\da_i\m\da_j\m(-\Delta)^{-1}
-{_1\over^3}\s\delta_{ij}\m)\m\s\CC\m)(0)\s\s x^i\m x^j\s+\s
{_1\over^{12\m D_1}}\s\s\CC(0)\s\s\vert x
\vert^2\s+\s\CO(L^{-1})\ .\hs{1cm}
\label{479}
\qqq
Thus, up to a constant and linear terms (the latter absent
for real \s$\CC\s$) with \s$\CC$-dependent, \s$L$-diverging
coefficients and a quadratic zero mode of \s$\CM_2^{\rm sc}\s$
with a \s$\CC$-dependent coefficient, we obtain in the
limit \s$L\to\infty\s$
the universal scaling solution \s$F_2^{\rm sc}(\Nx)\s=\s
{\epsilon\over 6\m D_1}\s\m|\Nx|^2\s$, \s in accordance
with the general result (\ref{limi}). Note that the separation
of the scaling solution from the zero mode of the
same homogeneity is ambiguous.
\vs 0.4cm

The $4$-point function in the stationary state
is given by Eqs.\s\s(\ref{pairi}) and (\ref{1chan})
for \s$M=2\s$, \s i.\s e. it may be written as a sum
of contributions of three $2$-particle channels:
\qq
<\prod\limits_{n=1}^4T(\Nx_n)>\s=\s
F_4(\Nx_1,\Nx_2,\Nx_3,\Nx_4)\s+\s
F_4(\Nx_1,\Nx_3,\Nx_2,\Nx_4)\s+\s
F_4(\Nx_1,\Nx_4,\Nx_2,\Nx_3)\hs{0.7cm}
\label{481}
\qqq
with the single channel function
\qq
\cr
F_4&=&\CM_4^{-1}\ (\m\CM_2^{-1}\otimes 1\s
+\s 1\otimes\CM_2^{-1})
\ \CC_L\otimes\CC_L\ .
\label{482}
\qqq
Working in the difference variables, we obtain
for \s$\kappa=0\s$
using the scaling operators (\ref{47}), (\ref{471})\s:
\qq
F_4(\Nx_1,\Nx_2,\Nx_3,\Nx_4)\s&=&\s{_1\over^{4\m D_1^2}}
\int\ee^{\m i\m(\Nk_1\cdot\Nx_1+\dots+\Nk_4\cdot\Nx_4)}
\ (\sum\limits_{n<n'}\Nk_n\cdot\Nk_{n'})^{-1}\cr
&&\cdot\s\s
(\m(\Nk_1\cdot\Nk_2)^{-1}\s+\s(\Nk_3\cdot\Nk_4)\m)^{-1}
\ \hat\CC_L(\Nk_1)\ \hat\CC_L(\Nk_3)\cr
&&\cdot\s\s(2\pi)^3\s
\delta(\Nk_1+\Nk_2)\s\s(2\pi)^3\s\delta(\Nk_3+\Nk_4)
\s\prod\limits_{n=1}^4{_{d^3\Nk_n}\over^{(2\pi)^3}}\cr
&=&\s{_1\over^{4\m D_1^2}}
\int\ee^{\m i\m(\Nk_1\cdot(\Nx_1-\Nx_2)
+\Nk_3\cdot(\Nx_3-\Nx_4)}
\ (\m\Nk_1^2+\Nk_3^2\m)^{-1}\s\s
(\m(\Nk_1^2)^{-1}+(\Nk_3^2)^{-1}\m)\cr
&&\hs{5.7cm}\cdot\s\s\hat\CC_L(\Nk_1)\ \hat\CC_L(\Nk_3)
\s\s{_{d^3\Nk_1}\over^{(2\pi)^3}}
\s\s{_{d^3\Nk_3}\over^{(2\pi)^3}}\cr
&=&\s F_2(\Nx_1-\Nx_2)\s\s F_2(\Nx_3-\Nx_4)\ ,
\qqq
see Eq.\s\s(\ref{2fc}). It follows from Eq.\s\s(\ref{481})
that for \s$\kappa=0\s$ the equal time \s$4$-point function
is expressed by the \s$2$-point function by the standard
Gaussian formula. This remains true for the higher point
functions. Let us consider the \s$6$-point function
\qq
<\prod\limits_{n=1}^6T(\Nx_n)>\s\ =\hs{-0.1cm}
\sum\limits_{\sigma\s\in\s\CS_6\s/\s\CS_3\times\CS_2^3}\hs{-0.3cm}
F_6(\Nx_{\sigma(1)},\dots,\Nx_{\sigma(6)})\ ,
\qqq
where \s$F_6\s$ is the contribution of a single
\s$2$-particle channel,
\qq
F_6\s=\s\CM_6^{-1}\sum\limits_{\sigma\in\CS_3}\sigma\m\left(
(\CM_4^{-1}\otimes 1_{2})\s\s(\CM_2^{-1}\otimes 1_{4})\right)
\s\CC_L\otimes\CC_L\ ,
\label{kanal}
\qqq
Explicitly, for \s$\kappa=0\s$,
\qq
F_6(\Nx_1,\dots,\Nx_6)\s&=&\s
{_1\over^{8\m D_1^3}}
\int\ee^{\m i\m(\Nk_1\cdot(\Nx_1-\Nx_2)
+\Nk_3\cdot(\Nx_3-\Nx_4)+\Nk_5\cdot(\Nx_5-\Nx_6)}
\ (\m\Nk_1^2+\Nk_3^2+\Nk_5^2\m)^{-1}\cr
&&\cdot\s\s(\m(\m\Nk_1^2+\Nk_3^2\m)^{-1}
+(\m\Nk_1^2+\Nk_5^2\m)^{-1}
+(\m\Nk_3^2+\Nk_5^2\m)^{-1}\m)\s\s
(\m(\Nk_1^2)^{-1}\cr
&&+(\Nk_3^2)^{-1}+(\Nk_5^2)^{-1}\m)
\s\s\hat\CC_L(\Nk_1)\ \hat\CC_L(\Nk_3)\
\hat\CC_L(\Nk_5)
\s\s{_{d^3\Nk_1}\over^{(2\pi)^3}}
\s\s{_{d^3\Nk_3}\over^{(2\pi)^3}}\s\s
{_{d^3\Nk_5}\over^{(2\pi)^3}}\cr\cr
&=&\s F_2(\Nx_1-\Nx_2)\s\s F_2(\Nx_3-\Nx_4)\s\s
F_2(\Nx_5-\Nx_6)\ .
\qqq
It should be clear how to extend this argument to
a general \s$2M$-point function.
\vs 0.3cm

If we want to separate in \s$F_{2M}\s$
the non-universal and the universal terms, we have
to extract from equation (\ref{seco}) more detailed
knowledge about the \s$2$-point function than that
given by the expansion (\ref{479}). Since \s$F_2\s$
starts with an \s$\CO(L^2)\s$ term, we have
to include into the expansion of, for example, \s$F_4\s$ all
terms of \s$F_2\s$ down to the order \s$\CO(L^{-2})\s$.
\s Since
\qq
F_2(\Nx)\ =\ {_1\over^{2\m D_1}}\sum\limits_{n=0}^4{_1\over^{n!}}
\s\m L^{2-n}
\s\s(\m\da_{i_1}\dots\da_{i_n}\s(-\Delta)^{-1}\s\CC\m)(0)
\s\s x^{i_1}\cdots x^{i_n}\ +\ \CO(L^{-3})\ ,
\qqq
\qq
F_4(\NX,\NY,\NZ)\ =\
{_1\over^{4\m D_1^2}}\sum\limits_{{m,n\geq 0}\atop{n+m\leq 4}}
{_1\over^{m!}}\s{_1\over^{n!}}\s\m L^{4-m-n}
\s\s(\m\da_{i_1}\dots\da_{i_m}\s\Delta^{-1}
\CC\m)(0)\hs{2cm}\cr
\cdot\m\s(\m\da_{j_1}\dots\da_{j_n}\s\Delta^{-1}\s\CC\m)(0)
\s\s X^{i_1}\cdots X^{i_m}\s\s Z^{j_1}\cdots Z^{j_n}\
+\ \CO(L^{-1})\ .
\qqq
For rotationally invariant \s$\CC\s$, \s this reduces to
\qq
F_4(\NX,\NY,\NZ)\ =\ {_1\over^{4\m D_1^2}}
\m\bigg(L^4\s\s(\Delta^{-1}\CC)(0)^2
\s+\s{_1\over^6}\s\m L^2\s\s\CC(0)\s\m(\Delta^{-1}\CC)(0)
\s\s(\m\vert\NX\vert^2+\m\vert\NZ\vert^2\m)\ \s\hs{0.5cm}\cr
-\s{_{1}\over^{144}}\s\m(\Delta\CC)(0)\s\s
(\Delta^{-1}\CC)(0)\s\s(\m\vert\NX\vert^4
+\m\vert\NZ\vert^4\m)\s+\s{_1\over^{36}}\s\s\CC(0)^2\s\m
\vert\NX\vert^2\m\vert\NZ\vert^2\bigg)\s+\s\CO(L^{-2})\ .
\hs{0.5cm}
\label{RoT}
\qqq
Note the presence of the universal Gaussian
contribution \s$F_2^{\rm sc}(\NX)\s F_2^{\rm sc}(\NZ)
={\epsilon^2\over 36\m D_1^2}\s\m|\NX|^2\s|\NZ|^2$
\s on the right hand side.
The rest of the terms are
non-universal zero modes of the operator
\qq
\CM_{2}^{\rm sc}\otimes\CM_{2}^{\rm sc}\s
(\CM_{2}^{\rm sc}\otimes 1+
1\otimes\CM_{2}^{\rm sc})^{-1}\s\CM_{4}^{\rm sc}
\qqq
equal to \s\s$\CM_{2}^{\rm sc}\otimes\CM_{2}^{\rm sc}\s=
\s4\m D_1^2\s\Delta_{\NX}\m\Delta_\NZ\s\s$
in the action on \s$\NY$-independent functions. That this
must be the case may
be easily seen by rewriting the relation (\ref{482})
as the equation
\qq
(\m\CM_{2}^{\rm sc}\otimes\CM_{2}^{\rm sc}\m)\s\s
(\CM_{2}^{\rm sc}\otimes 1+
1\otimes\CM_{2}^{\rm sc})^{-1}\s
\CM_{4}^{\rm sc}\ F_4\ =\ \CC_L\otimes\CC_L\ .
\label{for2}
\qqq
Note that the separation of the universal Gaussian
contribution \s$F_2^{\rm sc}(\NX)\s\s F_2^{\rm sc}(\NZ)\s$
from the terms with \s$m=n=2\s$
is ambiguous since there
are non-universal terms of the same homogeneity degree.
Nevertheless, the calculation for \s$\kappa=0\s$
confirms the general structure of the correlations
with the non-universal terms given by homogeneous
zero modes of differential operators.
\vs 0.7cm

\subsection{$4$-point and $6$-point
function for non-zero $\kappa\s$}
\vs 0.3cm

Let us consider the \s$4$-point function given
by Eqs.\s\s(\ref{481}) and (\ref{482}) for general
\s$\kappa\s$ between $0$ and $2$.
Recalling that \s$F_2=\CM_2^{-1}\m\CC_L\s$,
\s it may be convenient to view Eq.\s\s(\ref{482})
as the differential equation for \s$F_4\s\m$:
\qq
\CM_4\s\s F_4\s=\s F_2\otimes\CC_L\s+\s\CC_L\otimes F_2\ .
\label{m41}
\qqq
The connected part \s$F^c_4=F_4-F_2\otimes F_2\s$
of \s$F_4\s$ satisfies the equation
\qq
\CM_4\s\s F^c_4\s=\s\CL\s\m(\m F_2\otimes F_2\m)\ ,
\label{esseq}
\qqq
where \s$\CL\s\equiv\s(\m\CK)_{1,3}+
(\m\CK)_{1,4}+(\m\CK)_{2,3}+(\m\CK)_{2,4}\s$, \s see
Eq.\s\s(\ref{eqcon}).
\vs 0.3cm

First, let us study the \s$L$-dependence of \s$F_2\otimes
F_2\s$. \s Since, as follows from
Eq.\s\s(\ref{m1}) with \s$\nu=0\s$,
\qq
F_2(\Nx)\ =\
\gamma\s\s L^{2-\kappa}\s
-\s{_{\epsilon}\over^{3\m(2-\kappa)\m D_1}}
\s\s\vert\Nx\vert^{2-\kappa}\ +\ \CO(L^{-2})\ ,
\qqq
we infer that the product of \s$2$-point functions
\qq
F_2(\NX)\s\s F_2(\NZ)\ =\ \s\gamma^2\s\s L^{2(2-\kappa)}
\ +\ \gamma\s\s{_{\epsilon}\over^{3\m(2-\kappa)\m D_1}}\s\s
L^{2-\kappa}\s(\m\vert\NX\vert^{2-\kappa}\s+\s
\vert\NZ\vert^{2-\kappa}\m)\ \s\cr
+\ {_{\epsilon^2}\over^{9\m(2-\kappa)^2\m D_1^2}}
\s\s\vert\NX\vert^{2-\kappa}\s\vert\NZ\vert^{2-\kappa}\
+\ \CO(L^{-\kappa})\ .
\label{2raz2}
\qqq
Compare this expression with the similar expression
(\ref{RoT}) for \s$\kappa=0\s$. \s The term \s$\sim
\s(\m\vert\NX\vert^4
+\m\vert\NZ\vert^4\m)\s$ present there comes from the
\s$\CO(L^{-\kappa})\s$ terms of the \s$\kappa>0\s$ case.
\vs 0.3cm

Let us return to the connected contribution to \s$F_4\s$
as given by Eq.\s\s(\ref{esseq}).
In the action on translationally invariant functions,
\qq
\CM_4\ =\s
&-&\tilde D^{ij}(\NX)\s\m\da_{X^i}\s\da_{X^j}
\s-\s\tilde D^{ij}(\NY)\m\s\da_{Y^i}\s\da_{Y^j}
\s-\s\tilde D^{ij}(\NZ)\m\s\da_{Z^i}\s\da_{Z^j}\ \s\cr
&-&\left(\m\tilde D^{ij}(\NX+\NY)-
\tilde D^{ij}(\NX)-\tilde D^{ij}(\NY)
\right)\m\da_{X^i}\s\da_{Y^j}\ \s\cr
&-&\left(\m\tilde D^{ij}(\NY+\NZ)-\tilde
D^{ij}(\NY)-\tilde D^{ij}(\NZ)
\right)\m\da_{Y^i}\s\da_{Z^j}\ \s\cr
&-&\left(\tilde D^{ij}(\NX+\NY+\NZ)-\tilde D^{ij}(\NX+\NY)
-\tilde D^{ij}(\NY+\NZ)+\tilde D^{ij}(\NY)\right)
\s\da_{X^i}\s\da_{Z^j}\hs{0.9cm}
\qqq
for \s$\nu=0\s$. \s Note that
\qq
\CL\s\m(\m F_2\otimes F_2\m)(\NX,\NY,\NZ)\ =&&
\bigg(\tilde D^{ij}(\NX+\NY+\NZ)-\tilde D^{ij}(\NX+\NY)\cr
&&-\tilde D^{ij}(\NY+\NZ)
+\tilde D^{ij}(\NY)\bigg)\s\s(\da_i\m F_2)(\NX)\s\s
(\da_j\m F_2)(\NZ)\hs{0.5cm}
\label{495}
\qqq
has a well defined
limit when \s$L\to\infty\s$ since it involves only
the derivatives of \s$F_2\s$.
\s In particular
for \s$m=0\s$ and \s$\nu=0\s$, the \s$L\to \infty\s$ limit
of \s$\CL\s\m(\m F_2\otimes F_2\m)\s$ is equal to
\qq
{_{\epsilon^2}\over^{9\s(2-\kappa)^2\s D_1^2}}
\s\s\CL^{\rm sc}\s\s\vert\NX\vert^{2-\kappa}
\m\vert\NZ\vert^{2-\kappa}
\qqq
and is a homogeneous
(rotationally-invariant)
function of \s$\NX,\m\NY,\m\NZ\s$ of degree \s$2-\kappa\s$
whose explicit form may be easily found.
It is not difficult to write down a solution
\s${F^c_4}^{\s\rm sc}\s$
of Eq.\s\s(\ref{esseq})
for the limiting case. Indeed, for
\qq
{F^c_4}^{\s\rm sc}\ =\ {_{\epsilon^2}
\over^{6\m(2-\kappa)^2(5-\kappa)\m D_1^2}}
\s\s(\s\vert\NX\vert^{2(2-\kappa)}
+\m\vert\NZ\vert^{2(2-\kappa)}\m)
\s-\s{_{\epsilon^2}\over^{9\s(2-\kappa)^2\s D_1^2}}
\s\s\vert\NX\vert^{2-\kappa}
\m\vert\NZ\vert^{2-\kappa}\ ,\
\label{4scal}
\qqq
one obtains with the use of the decomposition
\m\s$\CM_4^{\rm sc}\s=
\s\CM_2^{\rm sc}\otimes1\s+\s1\otimes\CM_2^{\rm sc}
\s-\s\CL^{\rm sc}\s$ the required relation
\qq
(\m\CM_4^{\rm sc}\s\s F_4^{\rm sc}\m)(\NX,\NY,\NZ)\
=\ {_{\epsilon^2}\over^{9\s(2-\kappa)^2\m D_1^2}}\
\CL^{\rm sc}\ \vert\NX\vert^{2-\kappa}
\m\vert\NZ\vert^{2-\kappa}\ \
\qqq
since \s\s$\CM_2^{\rm sc}\otimes1+1\otimes\CM_2^{\rm sc}\s\s$
does not contribute to
the left hand side and \s\m$\CL^{\rm sc}\m\s$ annihilates
functions depending only on \s$\NX\s$ or only on \s$\NZ\s$.
By the same argument as for the two-point function, the
solution for finite but large \s$L\s$ should differ
from the universal scaling form by zero modes of
\s$\CM_4^{\rm sc}\s$ so that
\qq
F^c_4\ \s-\sum\limits_{0\leq\zeta_{4,n}\leq 2(2-\kappa)}
\hs{-0.25cm}L^{2(2-\kappa)
-\zeta_{4,n}}\sum\limits_m\gamma_{nm}\s\s F^c_{4,nm}
\ \ \
\smash{\mathop{\longrightarrow}
\limits_{L\to\infty}}\ \ \ {F^c_4}^{\s\rm sc}
\label{AsymP}
\qqq
where \s$F^c_{4,nm}\s$ are homogeneous zero modes
of \s$\CM_4^{\rm sc}\s$ of degree \s$\zeta_{4,n}\s$
and the non-universal coefficients \s$\gamma_{nm}\s$
depend on the source covariance \s$\CC\s$. \s In
fact, a more
complicated behavior with logarithmic terms
\qq
F^c_4\ \s-\sum\limits_{0\leq\zeta_{4,n}\leq 2(2-\kappa)}
\hs{-0.25cm}L^{2(2-\kappa)
-\zeta_{4,n}}\sum\limits_{m,k}\gamma_{nmk}\s\s
(\m\ln{(L\m f_{nm})}\m)^k\s\s F^c_{4,nm}
\ \ \
\smash{\mathop{\longrightarrow}
\limits_{L\to\infty}}\ \ \ {F^c_4}^{\s\rm sc}
\label{AsmP}
\qqq
is possible if \s$\CM_4\s$ has zero modes
\s$(\m\ln{f_{nm}}\m)^l\s\s F_{nm}\s$, \s$l\leq k\s$, \s for
\s$f_{nm}\s$ homogeneous functions of degree $-1$.
\vs 0.4cm

The main problem
which remains in the analysis of the universality
of the $4$-point functions is to find homogeneous
zero modes of \s$\CM_4^{\rm sc}\s$ with degree
between zero and \s$2(2-\kappa)\s$ and to decide
which ones appear in the relation (\ref{AsymP}).
The study of such zero modes seems to lie beyond
the scope of the techniques \cite{math} developed for
the singular elliptic operators of the type
we encounter here.
Note that in the full connected $4$-point function
only the zero modes with permutation symmetry
survive. If the sources are not only homogeneous but
also isotropic, only \s$SO(3)$-symmetric zero
modes may appear. In fact, one should expect
that such zero modes do enter into the expansions
(\ref{AsymP}) or (\ref{AsmP}) for generic \s$\CC\s$
whenever not forbidden by symmetries (which is
a somewhat tautological statement).
Existence of the zero-mode
contributions to the \s$4$-point function is
important from the practical point of view. If
we study a combination of \s$4$-point functions
(e.\s g. the \s$4^{\rm\m th}\s$ structure function)
and we fix the range of variation of distances between
the points, then the non-universal
terms with the smallest \s$\zeta^{\rm min}_{4,n}
<2(2-\kappa)\s$
dominate the behavior for large \s$L\s$ and we would see
the anomalous exponent \s$\zeta_{4,n}^{\rm min}\s$
rather than \s$2(2-\kappa)\s$ as governing the
distance behavior. The possible appearance of zero modes
of \s$\CM_4^{\rm sc}\s$ in the asymptotic
behavior of the \s$4$-point function would lead
to the picture of {\bf restricted
universality}, holding strictly
only for the \s$4$-point correlator
{\bf infrared-renormalized} by subtracting homogeneous terms
with non-universal coefficients typically divergent
when the integral scale \s$L\s$ grows to infinity.
In the presence of non-universal terms
with \s$\zeta_{4,n}=2(2-\kappa)\s$ with are
\s$L$-independent, the separation between the universal
and non-universal terms in the \s$4$-point function
would become ambiguous and the Eq.\s\s(\ref{4scal})
would involve an arbitrary choice.
\vs 0.3cm

Notice that
adding the connected and disconnected contributions
to \s$F_4\s$, \s one obtains (in the absence of
logarithmic terms)
\qq
F_4(\NX,\NY,\NZ)\s =\s
G(\NX)\m+\m G(\NZ)\m+\m\sum
\limits_{\zeta_{4,n}}L^{2(2-\kappa)-\zeta_{4,n}}
\sum\limits_m\gamma_{nm}\s\m F^c_{4,nm}
(\NX,\NY,\NZ)\s\m+\s o(L)\hs{0.95cm}
\qqq
for  \s\s$G(\Nx)\s=\s
\hf\s\gamma^2\s\s L^{2(2-\kappa)}
\ +\ \gamma\s\s{_{\epsilon}\over^{3\m(2-\kappa)\m D_1}}\s\s
L^{2-\kappa}\s\vert\Nx\vert^{2-\kappa}\s
+\s{_{\epsilon^2}\over^{12\m(2-\kappa)^2\m(5-\kappa)\m D_1^2}}
\s\s\vert\Nx\vert^{2(2-\kappa)}\s$. \s\m
Only \s$F^c_{4,mn}\s$
terms may contribute
to the structure function \s$S_4(r)\s$, \s as
all the terms dependent on
only one variable difference do not survive in \s$S_4\s$.
Hence the presence of the non-trivial contributions from
the (generally) non-universal terms \s$F^c_{4,nm}\s$
is a necessary condition for the very non-triviality of
\s$S_4\s$ for large \s$L\s$.
\vs 0.4cm

The \s$6$ point function may be treated similarly.
The connected part \s$F^c_6\s$
of the single $2$-particle channel
contribution \s$F_6\s$ is given by
\qq
F_6(\Nx_1,\dots,\Nx_6)&=&F^c_6(\Nx_1,\dots,\Nx_6)
\s+\s F^c_4(\Nx_1,\dots,\Nx_4)\s\s F_2(\Nx_5,\Nx_6)\cr\cr
&+& F^c_4(\Nx_1,\Nx_2,\Nx_5\Nx_6)\s\s F_2(\Nx_3,\Nx_4)
\s+\s F^c_4(\Nx_3,\dots,\Nx_6)\s\s F_2(\Nx_1,\Nx_2)\cr\cr
&+&\s F_2(\Nx_1,\Nx_2)\s\s F_2(\Nx_3,\Nx_4)
\s\s F_2(\Nx_5,\Nx_6)\ ,
\qqq
see Eq.\s\s(\ref{cOn}). It satisfies Eq.\s\s(\ref{eqcon})
which reads
\qq
\CM_6\s\m F^c_6\ =\ (\m(\CL_{1,3}\s
+\s\CL_{2,3}\m)\s(\m F^c_4(1,2)\s F_2(3)\m)\s+\s
(\m\CL_{1,2}\s+\s\CL_{2,3}\m)\s(\m(F^c_4(1,3)\s F_2(2))\ \s\s\cr\cr
+\s(\m\CL_{1,2}\s
+\s\CL_{1,3}\m)\s(\m F_2(1)\s F^c_4(2,3)\m)\ ,\hs{1cm}
\label{cOS}
\qqq
where the numbers in the paranthesis label
the $3$ consecutive pairs of points. In the case that
the non-universal non-constant zero modes do not appear
in the connected \s$4$-point function, the right hand side of
Eq.\s\s(\ref{cOS}) has a finite \s$L\to\infty\s$ limit
so that the relation of the type (\ref{AsymP}),
\qq
F^c_6\ \s-\sum\limits_{0\leq\zeta_{6,n}\leq 3(2-\kappa)}
\hs{-0.25cm}L^{3(2-\kappa)
-\zeta_{6,n}}\sum\limits_m\gamma_{nm}\s\s F^c_{6,nm}
\ \ \
\smash{\mathop{\longrightarrow}
\limits_{L\to\infty}}\ \ \ {F^c_6}^{\s\rm sc}\s,
\label{AsP}
\qqq
with \s$F^c_{6,mn}\s$ being homogeneous zero modes
of the degree \s$\zeta_{6,n}\s$ of \s$\CM_6^{\rm sc}\s$
should hold for \s$m=0=\nu\s$
(with eventual additional logarithmic terms,
as in (\ref{AsmP})\m)\m. \m As
we shall see in the next Section
such zero modes with rotational
and permutation symmetries
exist at least for small \s$\kappa\s$. \s Hence,
the non-constant non-universal terms should appear
if not already in the connected \s$4$-point function
then in the connected \s$6$-point function.
Whether they are visible in the corresponding
structure functions is a more difficult question
to which we do not have definite answer yet.
\vs 0.4cm

Let us end this section by stressing a difference
between the infrared renormalization
required by the correlation
functions of the PS and the renormalization
of infrared divergences occurring in massless models of
statistical mechanics or field theory. There the
divergences are an artifact of the perturbative expansion
and signal that the control parameters of the system are
off their critical values. The divergences may be removed by
finite shifts of the control
parameters which depend non-smoothly
on the coupling constants (hence the infinities in the
perturbative treatment). On the contrary,
in the PS model
of turbulent advection, the genuine divergences appear when
the size of the system tends to infinity and the correlators
become finite only after the subtraction of the
diverging terms. Similar picture appeared in Polyakov's
$2$-dimensional conformal turbulence \cite{Polya}
where it was argued that
extracting the conformal-invariant
part of the correlation functions
requires subtraction of non-universal polynomial terms in
the position space\footnote{K.G. thanks A. Polyakov for
discussion of this point}. We expect the infrared renormalization
of the type described above
to be the general feature of universality in turbulent systems
and the most important lesson to learn from the study
of the PS even if, at present, we are unable to
decide whether non-universal terms are present
in the structure functions, the most commonly studied correlators.
\vs 0.7cm

\subsection{$\kappa$-expansion}
\vskip 0.3cm

We may study the homogeneous zero modes of
the operator \s$\CM_{4}^{\rm sc}\s$ in perturbation
expansion in powers of \s$\kappa\s$.
\s We shall work here only to the first order in \s$\kappa\s$.
\s Eq.\s\s(\ref{td}) implies that (for \s$m=0\s$)
\qq
\tilde D^{ij}(\Nx)\s=\s 2\m D_1\s\delta^{ij}\s+\s
\kappa\s D_1\m\left(
2\s\delta^{ij}\s\ln\vert\Nx\vert\s-\s x^i\m
x^j\s\vert\Nx\vert^{-2}
\s+\s\delta^{ij}\right)\s
+\s\CO(\kappa^2)\ \m\s\cr
\equiv\s2\m D_1\s\delta^{ij}\s+\s
2\m\kappa\s D_1\s\m R^{ij}(\Nx)\s+\s\CO(\kappa^2)\ .
\qqq
As a result, to the first order in \s$\kappa\s$,
\qq
\CM_4^{\rm sc}&=&\CM^{\rm sc}_{4,0}\s-\s
2\m\kappa\s D_1\s\m R^{ij}(\NX)\s\m\da_{X^i}\s\da_{X^j}
\s-\s2\m\kappa\s D_1\s\m R^{ij}(\NY)\m\s\da_{Y^i}\s\da_{Y^j}
\s-\s2\m\kappa\s D_1\s\m R^{ij}(\NZ)\m\s\da_{Z^i}\s\da_{Z^j}\ \s\cr
&-&2\m\kappa\s D_1\left(R^{ij}(\NX+\NY)-
R^{ij}(\NX)-R^{ij}(\NY)\right)\m\da_{X^i}\s\da_{Y^j}\ \s\cr
&-&2\m\kappa\s D_1\left(R^{ij}(\NY+\NZ)-R^{ij}(\NY)-R^{ij}(\NZ)
\right)\m\da_{Y^i}\s\da_{Z^j}\ \s\cr
&-&2\m\kappa\s D_1\left(R^{ij}(\NX+\NY+\NZ)-R^{ij}(\NX+\NY)
-R^{ij}(\NY+\NZ)+R^{ij}(\NY)\right)
\s\da_{X^i}\s\da_{Z^j}\cr\cr
&\equiv&\CM^{\rm sc}_{4,0}\s+\s2\m\kappa\s D_1\s\s V_4\ ,
\qqq
where the subscript "$0$" refers to \s$\kappa=0\s$.
\s$\CM_4^{\rm sc}\s$ commutes with $3$-dimensional translations
and rotations and with the permutations of four points.
We shall search for its zero modes respecting these symmetries.
Note that
\qq
\CM_{4,0}^{\rm sc}\s=\s-2\m D_1\s(\m\Delta_{\tilde\NX}+\Delta
_{\tilde\NY}+\Delta_{\tilde\NZ})\ ,
\qqq
where
\qq
\tilde\NX\s=\s\NX\s,\ \ \ \tilde\NY\s=\s
\sqrt{2}(\NY+\hf\NX+\hf\NZ)\s,\ \ \ \tilde\NZ\s=\s\NZ\ .
\qqq
Denoting \s$R\s\equiv\s(\tilde\NX^2+\tilde\NY^2
+\tilde\NZ^2)^{1/2}\s$,
\s we obtain
\qq
\CM^{\rm sc}_{4,0}\s=\s-\m{_{2\m D_1}\over^{R^8}}\s\da_R\s
R^8\s\da_R\s+\s{_{2\m D_1}\over^{R^2}}\s\Phi
\label{formo}
\qqq
where \s$\Phi\s$ is the Laplacian on the sphere \s$S^8\s$
in the space
of \s$(\tilde\NX,\tilde\NY,\m\tilde\NZ)\s$. \s The
symmetric zero modes of the lowest
homogeneity of \s$\CM_{4,0}^{\rm sc}\s$
occur in degree zero (constants) and in degree \s$4\s$.
The latter, when expressed in terms of point-differences,
have the form
\qq
a\hs{-0.2cm}\sum\limits_{\{n,n'\}}\hs{-0.1cm}
(\Nx_n\hs{-0.05cm}-\hs{-0.05cm}\Nx_{n'})^4\s+\s b
\hs{-0.85cm}\sum\limits_{\{\{n,m\},\m\{n,m'\}\}}\hs{-0.8cm}
(\Nx_n\hs{-0.05cm}-\hs{-0.05cm}\Nx_{m})^2\m
(\Nx_n\hs{-0.05cm}-\hs{-0.05cm}\Nx_{m'})^2
\s+\s c\hs{-0.85cm}
\sum\limits_{\{\{n,n'\},\{m,m'\}\}}\hs{-0.8cm}
(\Nx_n\hs{-0.05cm}-\hs{-0.05cm}\Nx_{n'})^2\m
(\Nx_m\hs{-0.05cm}-\hs{-0.05cm}\Nx_{m'})^2\s,\hs{1cm}\label{zm2}
\qqq
where the pairs \s$\{n,n'\}\s$ and \s$\{m,m'\}\s$
are assumed different, as well as the pairs \s$\{n,m\}\s$
and \s$\{n,m'\}\s$ and where
\qq
20\m a\m+\m 28\m b\m+\m 6\m c\m=\m 0\ .
\label{abc}
\qqq
The constant survives as the
eigenvalue of \s$\CM_4^{\rm sc}\s$
for \s$\kappa\not=0\s$. \s The change
of the two independent zero
modes \s$R^4\m f_i\s$, \s$i=1,2\s$, \s of the form
(\ref{zm2}) may be traced by the standard
degenerate perturbation theory. This is done as follows.
Looking for a homogeneous
zero mode \s$R^{4+\kappa\lambda}\s(\m a_1\s f_1\s+\s
a_2\s f_2\s+\s\kappa\s f_3\m)\s$
with a homogeneous degree zero
function \s$f_3\s$ orthogonal
to \s$f_{1,2}\s$ in \s$L^2(S^{8})\s$, \s we obtain
in the linear order in \s$\kappa\s$
\qq
\CM_{4,0}^{\rm sc}\m
\left(\lambda\s R^4\s\ln{R}\s\m(a_1 f_1+a_2 f_2)
\s+\s R^4\s\m f_3\m\right)\s\m
+\m\s 2\m D_1\s\m V_4\s\m R^4\s(a_1 f_1+a_2 f_2)\s=\s0
\qqq
or, using the form (\ref{formo}) of \s$\CM_{4,0}^{\rm sc}\s$,
\qq
-15\s\lambda\m\s(a_1 f_1+a_2 f_2)\s-\s44\s f_3\s+\s\Phi\s f_3
\s+\s{_1\over^{R^2}}\s\m V_4\s R^4\s(a_1 f_1+a_2 f_2)\s=\s0\ .
\label{pe0}
\qqq
Upon taking the \s$L^2(S^8)\s$
scalar products with \s$f_{1,2}\s$, \s$f_3\s$ drops out
resulting in the relation
\qq
-15\s\lambda\s\sum\limits_{j=1,2}(\m f_i\m,\m f_j)\s\m a_j
\s+\s\sum\limits_{j=1,2}
(\m f_i\m,\s{_1\over^{R^2}}\s V_4\s R^4\s f_j)
\s\m a_j\s=\s0
\label{pe}
\qqq
Hence \s$\lambda\s$ has to solve the equation
\qq
\det\left(\s(\m f_i\m,\s{_1\over^{15\s R^2}}
\s V_4\s R^4\s f_j)\s-
\s\lambda\s\m(\m f_i\m,\s f_j)\right)\s=\ 0\ .
\label{dEt}
\qqq
It may be decided numerically whether
this equation has solutions \s$\lambda<-2\s$ so that
there exist homogeneous zero mode of \s$\CM_4^{\rm sc}\s$
with degree \s$\leq\s 4-2\kappa\s$ for small \s$\kappa\s$
which, multiplied by \s$\gamma\s\m L^{-(2+\lambda)\m\kappa}\s$,
\s should appear as non-universal terms in the
truncated \s$4$-point function. Given \s$\lambda\s$
and a non-zero solution \s$(a_1,a_2)\s$ of Eq.\s\s(\ref{pe}),
the \s$\CO(\kappa)\s$ contribution \s$f_3\s$ to the zero mode
should be computed from Eq.\s\s(\ref{pe0}).
\vs 0.3cm

If Eq.\s\s(\ref{dEt}) has one double root but
there is only one solution \s$(a_1,a_2)\s$
(up to normalization) of Eq.\s\s(\ref{pe}) then
one should look for an additional zero mode of
\s$\CM_4^{\rm sc}\s$ of the form
\qq
\kappa\s\s(\m\ln{R}\m)\s\s
R^{4+\kappa\lambda}\s(\m a_1\s f_1\s+\s
a_2\s f_2\s+\s\kappa\s f_3\m)\s+\s
R^{4+\kappa\lambda}\s(\m b_1\s f_1\s+\s
b_2\s f_2\s+\s\kappa\s f_4\m)\ .
\qqq
This leads in the first order to the relation
\qq
-15\s((a_1+\lambda\m b_1) f_1+(a_2+\lambda\m b_2)
 f_2)\m-\m44\s f_4\m+\m\Phi\s f_4
\m+\m{_1\over^{R^2}}\s V_4\s
R^4\s(b_1 f_1+b_2 f_2)\s=\s0\hs{0.8cm}
\qqq
which implies that
\qq
-15\m\sum\limits_{j=1,2}
(\m f_i\m,\m f_j)\s(\m a_j+\lambda\s b_j\m)
\s+\s\sum\limits_{j=1,2}
(\m f_i\m,\s{_1\over^{R^2}}\s V_4\s R^4\s f_j)
\s\s b_j\s=\s0\ .
\qqq
The latter equation has a non-zero
solution for \s$(b_1,b_2)\s$.
In this case, the logarithmic terms
\s\s$\sim\s(\m\ln{L}\m)\s L^{-(2+\lambda)\m\kappa}\s$
should appear in the asymptotic expansion of \s$F^c_4\s$
for \s$\kappa>0\s$ if \s$\lambda\leq -2\s$.
\vs 0.4cm

The zero modes of the scaling operator \s$\CM_{6}^{\rm sc}\s$
may be studied similarly by writing
\qq
\CM_6^{\rm sc}\s=\s\CM_{6,0}^{\rm sc}\s+
2\s\kappa\s D_1\m\s V_6\ ,
\qqq
and using the perturbation expansion. By a linear
change of variables, \s$\CM^{\rm sc}_{6,0}\s$
is equivalent to the Laplacian in \s$\NR^{15}\s$.
As for \s$\CM_{4,0}^{\rm sc}\s$, \s the homogeneous
zero modes of \s$\CM_{6,0}^{\rm sc}\s$ with the lowest
degree, symmetric under rotations and permutations,
are constants and the $4^{\m\rm th}$-order polynomials
(\ref{zm2}) satisfying this time
the relation
\qq
20\m a\m+\m 56\m b\m+\m 36\m c\m=\m 0\ .
\qqq
For \s$\kappa>0\s$ the constants are still
annihilated  by \s$\CM_{6,0}^{\rm sc}\s$ whereas
the two $4^{\m\rm th}$-order zero modes may change
slightly the homogeneity degree (one of them may also
pick a logarithm). As for small \s$\kappa\s$
the deformed degree will still be smaller than
\s$3(2-\kappa)\s$, \s  the deformed zero modes,
accompanied by a positive power of \s$L\s$ and a
non-universal coefficient, should appear in the
asymptotic expansion of the connected $6$-point
function discussed above.
\vs 1cm

\nsection{Conclusions}
\vskip 0.5cm

At the beginning of these lectures, we have sketched the
main idea of Kolmogorov's picture of universality
in the fully developed turbulence, related to the
energy cascade. We briefly described
how the turbulence problem may be formulated
in the language formally very close to that
of equilibrium statistical mechanics or
Euclidean field theory stressing, however,
the important differences between the two classes
of problems. The bulk of the lectures was devoted
to the discussion of a simple model of
turbulent phenomenon, the passive scalar (PS),
which we treated as a test ground for the
universality ideas. The exact solution for the model
in terms of singular elliptic many-body operators
showed non-universal diffusive decay of correlations for
deterministic initial data but a universal super-diffusive
decay for random homogeneous initial data. In the presence
of random sources injecting the energy at small wavenumbers,
the steady state exhibited an inertial range
with the energy cascade and
universal scaling in the connected equal-time
two-point function of the scalar. We have argued that
the higher-point equal-time connected functions
are expected to show only restricted universality
with the terms dominating for large size systems
characterized by universal
anomalous exponents and non-universal
amplitudes. We have shown that such terms should appear
in the $6$-point function if not in the $4$-point one,
at least for small velocity exponent \s$\kappa\s$.
Further work along well traced direction is required
to decide on this question as well as on the question
what terms dominate the respective structure functions.
One of the open problems is to relate the terms
with anomalous exponents to the physical idea
of intermittency. One should also understand what role,
if any, is played in the model by the volume preserving
diffeomorphisms.
Upon infrared renormalization which subtracts the
non-universal terms, the equal-time correlators
of the scalar should become finite
and universal (with normal scaling) in the limit
when molecular diffusivity
\s$\nu\to0\s$ and the integral scale \s$L\to\infty\s$.
In particular, no \s$\nu$-dependent subtractions
are needed in the steady-state equal-time correlators
of the scalar and their short-distance behavior is
easy to analyze \cite{Russ}\cite{Proc}. This behavior
determines what ultraviolet (short distance)
renormalization is required by the composite operators
involving the gradients of \s$T\s$.
The basic role of \s$\nu\s$ is to assure the convergence
to the steady state by providing the mechanism which
removes on small scales the energy injected by the
sources at large distances.
\vs 0.3cm

We expect the main
features of the behavior of the PS
described above to be present also in other systems
exhibiting a fully developed turbulence, including
the Navier-Stokes flow. Still
much remains to be done even for the PS.
One important step could be a complete perturbative
analysis with the above picture of restricted
universality established order by order.
The \s$\kappa\s$ expansion discussed above
is a possibility but already the analysis
of the lowest terms proved to be quite complicated
\cite{my}.
Other possibilities are the quasi-Lagrangian
perturbation expansion or the expansion proposed
in \cite{Russ}. Ultimately, if no way to
compute exactly the non-universal terms in the correlators
is found, a multiscale analysis employing perturbative
arguments in conjunction with a renormalization group
type analysis should provide a tool to fully control
the behavior of correlation functions. Developing
such methods for the PS may be a first
step towards a complete theory of fully developed
turbulence. More than half-century after Kolmogorov's
work, despite further progress, a
fully developed understanding of the fully
developed turbulence
remains a major challenge.
\vs 1cm


\begin{thebibliography}{bib}
\bibitem{DoGr6}
Phase transitions and critical phenomena. Vol. 6,
eds. C. Doomb, M. S. Green. Academic Press, London 1976

\bibitem{DShK}
R. L. Dobrushin, R. Koteck\'{y}, S. Shlosman:
The Wulff construction: a global shape from local
interactions. AMS, Translations of Mathematical Monographs
104, Providence, Rhode Island 1992

\bibitem{FMRT}
J. Feldman, E. Trubowitz: {\it The Flow of an Electron-Phonon
System to the Superconducting State}. Helv. Phys. Acta {\bf 64}
(1991), 213-357

\bibitem{Fr}
J. Fr\"{o}hlich: contribution to this volume

\bibitem{Feig}
M. J. Feigenbaum: {\it Qualitative Universality for a Class
of Nonlinear Transformations}. J. Stat. Phys. {\bf 19}
(1978), 25-52

\bibitem{BrKup}
J. Bricmont, A. Kupiainen: {\it
Renormalizing Partial Differential Equations}.
Commun. Pure. Appl. Math. {\bf 47} (1994), 893-922

\bibitem{WilKo}
K. G. Wilson, J. Kogut: {\it The Renormalization Group
and the $\epsilon$-Expansion}. Phys. Rep. {\bf 12} (1974), 75-200

\bibitem{Wil}
K. G. Wilson: {\it The Renormalization
Group and Critical Phenomena}.
Rev. Mod. Phys. {\bf 55} (1983), 583-600

\bibitem{K41}
A. N. Kolmogorov: {\it The Local Structure of Turbulence
in Incompressible Viscous Fluid for Very Large Reynolds'
Numbers}. C. R. Acad. Sci. URSS {\bf 30} (1941), 301--305

\bibitem{Obu}
A. M. Obukhov: {\it Structure of the Temperature
Field in a Turbulent Flow}. Izv. Akad. Nauk SSSR, Geogr.
Geofiz. {\bf 13} (1949), 58-69

\bibitem{Kr94}
R. H. Kraichnan: {\it Anomalous Scaling of a Randomly Advected
Passive Scalar}. Phys. Rev. Lett. {\bf 72} (1994), 1016-1019

\bibitem{LPF}
V. S. L'vov, I. Procaccia, A. Fairhall: {\it Anomalous
Scaling in Fluid Dynamics: the Case of Passive Scalar}.
Phys. Rev. {\bf E 50} (1994), 4684-4704

\bibitem{Majda}
A. Majda: {\it Explicit Inertial Range Renormalization
Theory in a Model for Turbulent Diffusion}. J. Stat. Phys.
{\bf 73} (1993), 515-542

\bibitem{Russ}
M. Chertkov, G. Falkovich, I. Kolokolov, V. Lebedev:
{\it Normal and Anomalous Scaling of the Forth-Order
Correlation Function of a Randomly Advected Passive Scalar}
Weizmann Institute preprint, chao-dyn/95030001

\bibitem{Proc}
A. L. Fairhall, O. Gat, V. L'vov, I. Procaccia: {\it Anomalous
Scaling in a Model of Passive Scalar Advection: Exact Results}.
Weizmann Institute preprint 1995

\bibitem{Kra}
R. H. Kraichnan, V. Yakhot, S. Chen:
{\it Scaling Relations for a Randomly Advected Passive
Scalar Field}, preprint 1995, submitted to Phys. Rev. Lett.

\bibitem{Rich}
L. F. Richardson: Weather prediction by numerical process.
Cambridge University Press, Cambridge 1922

\bibitem{Frisch}
U. Frisch: Turbulence: the legacy of A. N. Kolmogorov.
Cambridge University Press. Cambridge 1995, to appear

\bibitem{MSR}
P. C. Martin, E. D. Siggia, H. A. Rose: {\it Statistical Dynamics
of Classical Systems}. Phys. Rev. {\bf A 8} (1973), 423-437
\bibitem{Polya}
A. M. Polyakov: {\it The Theory of Turbulence in Two Dimensions}.
Princeton University preprint, hep-th/9212145

\bibitem{BerlLv}
V. I. Belinicher, V. S. L'vov: {\it A Scale Invariant Theory
of Fully Developed Hydrodynamic Turbulence}. Sov. Phys. JETP
{\bf 66} (1987), 303-313

\bibitem{Lv}
V. S. L'vov: {\it Scale Invariant Theory of Fully
Developed Hydrodynamic Turbulence - Hamiltonian Approach}.
Phys. Rep. {\bf 207} (1991), 1-47

\bibitem{LvProc}
V. S. L'vov, I. Procaccia: {\it Exact Resummation of the Theory
of Hydrodynamic Turbulence. I. The Ball of Locality
and Normal Scaling}. Weizmann Institute preprint (1995),
submitted to Phys. Rev. {\bf E}

\bibitem{GlSh}
J. Glimm, D. H. Sharp: {\it A Random Field Model for
Anomalous Diffusion in Heterogeneous Porous Media}.
J. Stat. Phys. {\bf 62} (1991), 415-424

\bibitem{Kraich}
R. H. Kraichnan: {\it Small-Scale Structure of a Scalar
Field Convected by Turbulence}.
Phys. Fluids {\bf 11} (1968), 945-963

\bibitem{my}
K. Gaw\c{e}dzki, A. Kupiainen: in preparation

\bibitem{math}
C. E. Guti\'{r}errez, G. S. Nelson: {\it Bounds
for the Fundamental Solution of Degenerate
Parabolic Equations}. Commun. Partial Diff.
Eq. {\bf 13} (1988), 635-649

\end{thebibliography}
\end{document}